\documentclass[showpacs,showkeys,11pt,
preprint,preprintnumbers,nofootinbib,
groupedaddress,superscriptaddress,amsmath,amssymb]{revtex4}
\usepackage{amsfonts}
\usepackage{amsmath}
\usepackage{graphics}
\usepackage[export]{adjustbox}
\usepackage{epsfig}
\usepackage{url}
\usepackage{multirow}
\usepackage{feynmp}

\newcommand {\be}{\begin{equation}}
\newcommand {\ee}{\end{equation}}
\newcommand {\ba}{\begin{eqnarray}}
\newcommand {\ea}{\end{eqnarray}}

\begin{document}
\title{Correlation between the scaling factor of the yukawa coupling and cross section for the $e^+ e^- \rightarrow hhf\overline{f}$ ($f\neq t$) in type-I 2HDM}

\pacs{12.60.Fr, 
      14.80.Fd  
}
\keywords{Charged Higgs, MSSM, LHC}
\author{Ijaz Ahmed}
\author{Sehrish Gul}
\email{Ijaz.ahmed@cern.ch}
\affiliation{Riphah International University, Sector I-14, Hajj Complex, Islamabad Pakistan}
\author{Taimoor Khurshid}
\email{taimoor.khurshid@iiu.edu.pk}
\affiliation{International Islamic University, H-10, Islamabad}


\begin{abstract}
The objective of this study is to correlate the scaling factor of the Standard Model (SM) like Higgs boson and the cross section ratio of the process $e^+ e^- \rightarrow hhf\overline{f}$ where $f\neq t$, normalized to SM predictions in the type I of the Two Higgs Doublet Model. All calculations have been performed at $\sqrt{s}=500$ GeV and $1 \leq \tan{\beta} \leq 30$ for masses $m_H = m_A = m_{H^\pm} = 300GeV$ and $m_H=300$ GeV, $m_A=m_{H^\pm}=500$ GeV. The working scenario is by taking without alignment limit, that is $s_{\beta-\alpha}= 0.98$ and $s_{\beta-\alpha}= 0.99,$ $0.995$, which gives the enhancement in the cross section, particularly a few times greater than the predictions of the SM due to resonant-impacts of the additional heavy neutral Higgs bosons. This shows that enhancement in cross section occurs on leaving the alignment i.e., $s_{\beta-\alpha}=1$, at which all the higgs that couple to vector bosons and fermions have the same values as in SM at tree level. A large value of enhancement factor is obtained at $s_{\beta-\alpha}= 0.98$ compared to $s_{\beta-\alpha}= 0.99,$ $0 .995$. Furthermore, the decrease in the enhancement factor is observed for the case when $m_H=300$ GeV, $m_A=m_{H^\pm}=500$ GeV. The behavior of the scaling factor with $\tan{\beta}$ is also studied, which shows that for large values of $\tan\beta$, the scaling factor becomes equal to $s_{\beta-\alpha}$. Finally a convincing correlation is achieved by taking into account, the experimental and theoretical  constraints e.g, perturbative unitarity, vacuum stability and electroweak oblique parameters.  
\end{abstract}
\maketitle
\section{Introduction}
The Higgs boson was observed  first time in the CMS  and ATLAS experiments \cite{lab1, lab2}, since then it is the most mysterious particle for physicists among all the known particles till today. The comprehension of its unique properties and interaction is essential for understanding the Standard Model. Our cognition about the fundamental particles and their interactions will enhance if there occurs some deviation from the SM predictions. Having this thought higgs exploration can be done in a more effective manner by using a $e^+ e^-$ collider having high luminosity and energy. The generation of double Higgs is so far not discovered through LHC \cite{lab5} experimentation. The SM predicts that a Higgs boson pair can be generated via the interaction of the Higgs field with itself. \\
In this work the  production cross section of two higgs  at $e^+ e^-$ collider using the THDM \cite{lab9} are studied. The THDM can be checked at the collider projects by the two methods, one is to directly search the extra Higgs bosons for example charged Higgs boson and another is to indirectly search the deviations in observed Higgs boson (h) characteristics as expected by the SM. Disclosure of extra Higgs bosons will give a direct confirmation of the THDM but no such new particles are hitherto observed in  LHC. Therefore, obtaining deviations in the observed Higgs boson characteristics is the best way. The  production cross section of double higgs boson is studied for both the cases i.e with as well as without alignment at the LHC \cite{lab5} . Here we concider the later on case i.e without the alignment limit  because it can give significant modification in the cross section. The objective of this study is to discover the correlation among the  higgs boson couplings deviation and enhancement of cross section. They are anticipated to be correlated with one another, as h coupling deviation emerges in without the alignment limit case where the double higgs boson production process may be mediated by  additional neutral higgs bosons and so may provide substantial enhancement of the cross section \cite{lab8}. This study will explain that how sizable enhancement may be achieved by taking into consideration without the alignment case. The paper is organized in such a way that a concise introduction of THDM is provided in second section. In third section  theoretical as well as experimental constraints are given. In section 4 general characteristics of  Higgs boson pair generation in $e^+ e^-$ colliders are discussed. After this a mechanism is given for the calculation of enhancement factor and scaling factor. In section 5 a complete numerical analysis is accomplished. Finally conclusion is given in section 6.

\section{Two Higgs Doublet Model 2HDM}
Although the SM successfully explained most of the smallest scale physics phenomenon, but there are some problems within it for example neutrino oscillations \cite{lab3}, dark matter, dark energy \cite{lab4}, gravity etc. These may be resolved only by looking beyond SM. Standard Models's simplest extension is THDM. THDM has total five Higgs particles. These include two lighter higgs which are CP even, H (heavy higgs), the pseudoscalar being odd in CP transformation and two more Higgs bosons which are charged in nature \cite{lab9}. Moreover THDM has six physical parameters which are masses of four Higgs ($m_h,m_H,m_A,m_{H^\pm}$), the $\tan\beta$ which is the ratio of  two vacuum expectation values(VEV) and mixing angle $\alpha$ that diagonalizes the mass matrix of h and H. There are two particular cases i.e., at $\cos(\beta-\alpha) \rightarrow 0$ or $\sin(\beta-\alpha) \rightarrow 1$, when lighter Higgs(h) acts as the SM Higgs boson. Another case is when $\sin(\beta-\alpha) \rightarrow 0$ or $\cos(\beta-\alpha) \rightarrow 1$, the heavy Higgs(H) exhibits couplings like the SM. Here $\alpha$ and $\beta$ denote the mixing angles. These alignment limits play a significant role in deciding which Higgs among h or H is the discovered Higgs, as the observed Higgs is also CP even. There exist many motivations for use of THDM, one important motivation is the Higgs sector as SM does not give any fundamental ground to presume that the Higgs sector has only one higgs doublet. Both up and down quarks can not get mass simultaneously from one Higgs doublet. So there should be two doublets, one for giving mass to up quarks and another for down quarks. Another significant motivation for THDM is the hierarchy problem of yukawa couplings i.e the ratio of two third generation quarks masses i.e the top and bottom quark masses $m_t/m_b \approx 174/5 \approx 35$. According to SM both quarks obtain mass from the similar Higgs doublet, resulting non natural hierarchy among their related Yukawa couplings. But this problem can be solved by concidering two higgs doublet model i.e one doublet give mass to up quark while other give mass to bottom. \\
Depending on the couplings of fermions to $\Phi_1$ and $\Phi_2$, THDM has four types, labelled as Type-I, II, X and Y. In Type-I, one doublet couples with both leptons and quarks $\Phi_2$. In Type-II up like quarks (up, charm, top) couple $\Phi_2$ and down like quarks (down, strange, bottom) couple to $\Phi_1$. Both Type-II THDM and MSSM have similar Yukawa couplings. Therefore, MSSM is a specific type of THDM. It is noticeable that in Type-I and II, THDM, both down like quarks and leptons exhibit interactions with the same Higgs doublet. In Type-X all quarks have coupling with $\Phi_2$ whereas the leptons interact with $\Phi_1$. As the leptons have specific interactions as compared to the up like and down like quarks, so this type of THDM is named as Lepton Specific.Up like quarks interact  with $\Phi_2$ in Type-Y, down type quarks interact with $\Phi_1$ but the  Higgs doublet which interact with up like quarks also interact with leptons  \cite{lab9}. Therefore, this type is also named as Flipped. Conventionally the up like quarks interact to $\Phi_2$ doublet.\\

\subsection{Yukawa Coupling}
As explained in the above section, there are different modes of coupling of THDM WITH the SM fermions. Here we considered the Type-I THDM. So Type-1  Yukawa Lagrangian is\\
\begin{equation}
   \mathcal{L}_Y=Y_d {\overline{Q_L}} {\Phi_2} d_R + Y_u {\overline{Q_L}} \widetilde{\Phi}_2 u_R + Y_e {\overline{L_L}} {\Phi_2} e_R + h.c. 
   \label{eq:2.1}
\end{equation}\\

Here $\overline{Q_L}$ and $\overline{L_L}$ represent the left-handed lepton and quark doublets, $e_R$, $d_R$ and $u_R$ and indicate right-handed down-type quark, up-type quark and lepton singlets respectively, $Y_u$, $Y_d$ and $Y_e$ are the correspondent Yukawa coupling matrices and $\widetilde{\Phi}_2=i \sigma_2 \Phi^*$ (where $\sigma_2$ denote the Pauli matrix). When the weak eigenstates of $\Phi_2$ are expressed in physical form, then the above equation takes the form:

\begin{center}
\begin{equation}
 \begin{aligned}
 -\mathcal{L}_Y= \sum_{\psi=u,d,l} \biggl(\frac{m_{\psi}}{\nu} {\kappa_{\psi}^h} \overline{\psi} {\psi} h^0 + \frac{m_{\psi}}{\nu} \kappa_{\psi}^H \overline{\psi} {\psi} H^0 -i \frac{m_{\psi}}{\nu} \kappa_{\psi}^A \overline{\psi} {\gamma_5} {\psi} A^0 \biggr) + \\
  \biggl(\frac{V_{ud}}{\sqrt{2}\nu} {\overline{u}} (m_u {\kappa_{u}^A P_L} + m_d {\kappa_{d}^A P_R})dH^+ + \frac{m_l {\kappa_{l}^A}}{\sqrt{2}\nu} {\overline{\nu}_L l_R H^+} + h.c.\biggr). 
 \label{eq:2.2}
 \end{aligned}
 \end{equation}
\end{center}

where $\kappa_{i}^s$ denote the Yukawa couplings in the THDM whose values are given in the table \ref{table:1}.

\begin{table}[h!]
  \begin{center}
    \begin{tabular}{|c|c|c|c|c|c|c|c|c|} 
      \hline 
	${\kappa_{u}^h}$ & ${\kappa_{d}^h}$ & ${\kappa_{l}^h}$ & ${\kappa_{u}^H}$ &${\kappa_{d}^H}$ & ${\kappa_{l}^H}$& ${\kappa_{u}^A}$& ${\kappa_{d}^A}$&${\kappa_{l}^A}$ \\
	\hline
	$c_{\alpha}/ s_{\beta}$ & $c_{\alpha}/ s_{\beta}$  & $c_{\alpha}/ s_{\beta}$& $s_{\alpha}/ s_{\beta}$ &$s_{\alpha}/ s_{\beta}$ & $s_{\alpha}/ s_{\beta}$& $c_{\beta}/ s_{\beta}$&-$c_{\beta}/ s_{\beta}$&-$c_{\beta}/ s_{\beta}$ \\
	\hline
    \end{tabular}
    \caption{Yukawa Couplings in the THDM Type-I.}
    \label{table:1}
  \end{center}
\end{table}

The Yukawa interactions for third generation fermions in the Higgs basis \cite{lab8} has the form

\begin{center}
\begin{equation}
\mathcal{L}_Y=-{\overline{Q}_{L}^3} \frac{\sqrt{2}m_t}{\nu} ({\widetilde{\Phi}}+ {\xi_t \widetilde{\Phi}^{\prime}})t_R -{\overline{Q}_{L}^3} \frac{\sqrt{2}m_b}{\nu} ({\Phi}+ {\xi_b \Phi}^{\prime})b_R  -{\overline{L}_{L}^3} \frac{\sqrt{2}m_{\tau}}{\nu} ({\Phi}+ {\xi_{\tau} {\Phi}^{\prime}})\tau_R + h.c 
 \label{eq:2.3}
 \end{equation}
\end{center}

where $ \widetilde{\Phi}=i {\tau_2} {\Phi^*}$ and ${\widetilde{\Phi}^{\prime}}=i{\tau_2} {\Phi^{\prime *}}$. Here h.c denotes the hermitian
conjugate of the terms
\begin{center}
\begin{equation}
-{\overline{Q}_{L}^3} \frac{\sqrt{2}m_t}{\nu} ({\widetilde{\Phi}}+ {\xi_t \widetilde{\Phi}^{\prime}})t_R -{\overline{Q}_{L}^3} \frac{\sqrt{2}m_b}{\nu} ({\Phi}+ {\xi_b \Phi}^{\prime})b_R -{\overline{L}_{L}^3} \frac{\sqrt{2}m_{\tau}}{\nu} ({\Phi}+ {\xi_{\tau} {\Phi}^{\prime}})\tau_R
 \end{equation}
\end{center}
Where the factors $\xi_b$ and $\xi_{\tau}$ represent the choice of different types of Yukawa interaction, shown in table \ref{table:2}.\\

One thing is important to mention here that $\xi_t = \cot\beta$ remains same for all Yukawa interaction. Fermions and Higgs boson interaction terms \cite{lab8} can be extricated as

\begin{center}
\begin{equation}
 \begin{aligned}
\mathcal{L}_Y=-\sum_{f=t,b,\tau} \frac{m_f}{\nu} \overline{f} [(s_{\beta-\alpha} + \xi_f c_{\beta-\alpha})h + (c_{\beta-\alpha}- \xi_f s_{\beta-\alpha})  H - 2i I_f \xi_f \gamma_5 A]f  - \\ \frac{\sqrt{2}}{\nu} \overline{t} (m_b P_R - m_t P_L)H^+b  - \frac{\sqrt{2}}{\nu} \overline{\nu}_{\tau} m_{\tau} P_R H^+ \tau + h.c. \label{eq:2.4}
 \end{aligned}
 \end{equation}
\end{center}
Here, $I_t(I_{b,\tau}) = 1/2 (-1/2)$ and $P_{R,L} = \frac{1}{2}(1 \pm \gamma_5)$ represents the chirality projection operators.

\ifx
\begin{table}[h!]
  \begin{center}
\begin{tabular*}{80mm}{@{\extracolsep{\fill}}|c|c|c|}
\hline 
\textbf{Model} & $\mathbf{\xi_b}$ & $\mathbf{\xi_\tau}$ \\
\hline
Type-I &\hphantom{0}$\cot \beta$  & \hphantom{0}$\cot \beta$  \\
Type-II& $-\tan \beta$ & $-\tan \beta$ \\
Type-X & \hphantom{0}$\cot \beta$  & $-\tan \beta$ \\
Type-Y& $-\tan \beta$ & \hphantom{0}$\cot \beta$\\
\hline
\end{tabular*}
\caption{ \label{table:2}  $\xi_b$ and $\xi_\tau$ values for the four types of THDM.}
  \end{center}
\end{table}
\fi

\begin{table}[h]
\begin{tabular}{|c|c|c|}
\hline 
\textbf{Model} & $\mathbf{\xi_b}$ & $\mathbf{\xi_\tau}$ \\
\hline
Type-I &\hphantom{0}$\cot \beta$  & \hphantom{0}$\cot \beta$  \\
Type-II& $-\tan \beta$ & $-\tan \beta$ \\
Type-X & \hphantom{0}$\cot \beta$  & $-\tan \beta$ \\
Type-Y& $-\tan \beta$ & \hphantom{0}$\cot \beta$\\
\hline
\end{tabular}
\caption{\label{table:2}  $\xi_b$ and $\xi_\tau$ values for the four types of THDM.}
\end{table}

\subsection{The Higgs Potential}
This potential is the region that ascertains the Soft Symmetry Breaking (SSB) structure as well as the Higgs masses, mass eigenstates and self interactions. The Higgs potential depends on $\Phi_1$ and $\Phi_2$ and several different mixing parameters, its most simplified form is specified in the equation (\ref{eq:2.5}).

\begin{center}
\begin{equation}
 \begin{aligned}
V_{H} = m_{11}^2 \Phi_{I}^{\dagger} \Phi_{I}  + m_{22}^2 \Phi_{II}^{\dagger} \Phi_{II} - [m_{12}^2 \Phi_{I}^{\dagger} \Phi_{II} + h.c]  + \frac{\lambda_1}{2} (\Phi_{I}^{\dagger} \Phi_{I})^2 +  \frac{\lambda_2}{2}(\Phi_{II}^{\dagger} \Phi_{II})^2  \\ + \lambda_3(\Phi_{I}^{\dagger} \Phi_{I})^2(\Phi_{II}^{\dagger} \Phi_{II})^2  + \lambda_4 |\Phi_{I}^{\dagger} \Phi_{II}|^2 +\biggr[\frac{\lambda_5}{2} (\Phi_{I}^{\dagger} \Phi_{II})^2  +  \lambda_6 (\Phi_{I}^{\dagger} \Phi_{I})(\Phi_{I}^{\dagger} \Phi_{II}) \\ + \lambda_7 (\Phi_{II}^{\dagger} \Phi_{II})(\Phi_{I}^{\dagger} \Phi_{II})+h.c \biggr] \label{eq:2.5}
 \end{aligned}
 \end{equation}
\end{center}

Here $\Phi_{I}$ and $\Phi_{II}$ represent two doublets $\Phi_1$ and $\Phi_2$. In equation, first h.c. denotes the hermitian conjugate of $m^{2}_{12} \Phi^{\dagger}_{I}\Phi_{II}$ term and the final h.c. refers to the last three terms enclosed in the bracket. $m^{2}_{11,22,12}$ denote the mass mixing parameters. The parameters $m^{2}_{11,22}$, $\lambda_{1-4}$ are real, where as $m^{2}_{12}$ and $\lambda_{5-7}$ are in general complex. Therefore, there are total fourteen parameters in Higgs potential given in the Eq. (\ref{eq:2.5}), in which six are real parameters and the leftover eight are complex. To suppress Flavor Changing Neutral Currents (FCNC), $Z_2$ symmetry is set for the Higgs potential given in the equation (\ref{eq:2.5}), $Z_2$ symmetry is: $\Phi_1 \rightarrow \Phi_1$ and $\Phi_2 \rightarrow -\Phi_2$. The Higgs potential obeys the conditions of $Z_2$ symmetry when $m^{2}_{12} = \lambda_6 = \lambda_7 = 0 $ in equation (\ref{eq:2.5}). We assume that $m^{2}_{12}, \lambda_5$ are real by presuming that the CP symmetry remain constant under such assumptions. Another assumption is $\lambda_6 = \lambda_7 = 0$ but $m^{2}_{12} \neq0$. Under such assumptions $Z_2$ symmetry "softly breaks". The parameters $m^{2}_{11}$ and $m^{2}_{22}$ can be eliminated by applying tadpole conditions. So, there left eight parameters, six in the potential and two vacuum expectation values which are
\begin{equation}
m_{H^{\pm}} , m_A , m_H , m_h, M^2 , \tan\beta , \nu , \alpha \label{eq:2.6}
\end{equation}
here $ M^2 \equiv m_{12}^2/(s_{\beta}c_{\beta})$ is the soft breaking parameter and $m_h= 125 GeV$ and $\nu= 246 GeV$.

\section{Theoretical and Experimental Constraints}
There are number of theoretical constraints, like perturbativity, vacuum stability \cite{lab10}, perturbative unitarity \cite{lab11} and experimental limits obtained from LEP \cite{lab6}, LHC and tevatron \cite{lab5} experimentation which constrained the 2HDM parameters. It is noticeable that  perturbativity, vacuum stability, unitarity, also S, T and U constraints are implemented in 2HDMC public code \cite{lab12}.

\section{Double Higgs Boson Production}
\begin{figure}[h]
    \centering
    \includegraphics[width=12cm]{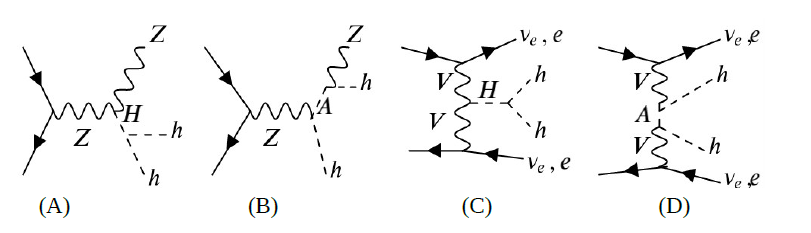}
    \caption{The double Higgs boson generation process in the electron positron collision in the SM, (here V= W or Z)}
    \label{fig:1}
\end{figure}

\begin{figure}[h]
    \centering
    \includegraphics[width=12cm]{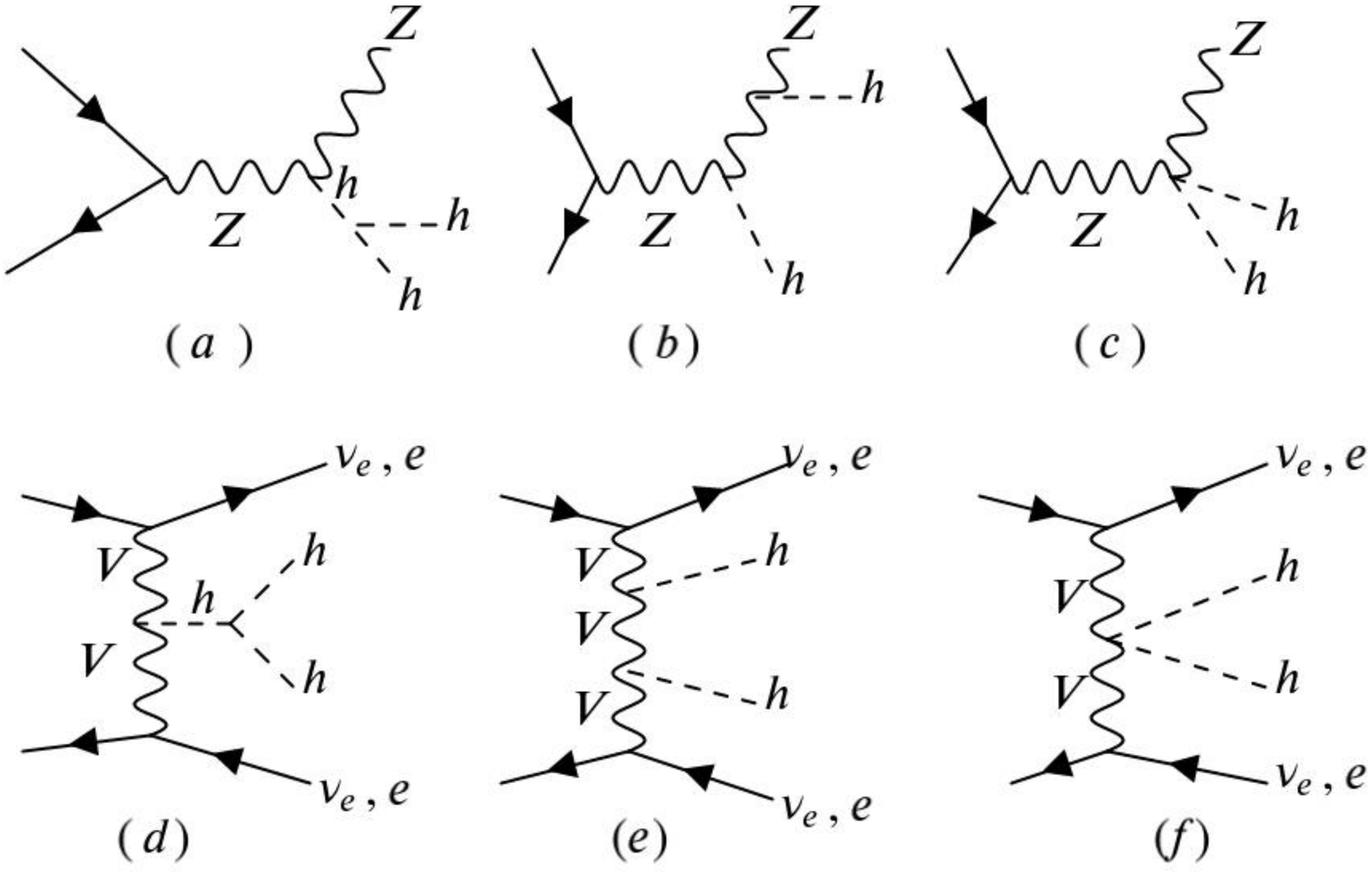}
    \caption{The double Higgs boson generation process in the electron positron collision appear in THDM, (where V= W or Z)}
    \label{fig:2}
\end{figure}

Here some production processes of double Higgs boson and differences of production cross section in THDM and SM are discussed. There are two major processes for Higgs pair generation at $e^+ e^-$ collider: double Higgs Strahlung process $e^+ e^- \rightarrow hhZ$ \cite{lab13} and W fusion process $e^+ e^- \rightarrow hh \nu_e \overline{\nu}_e$ \cite{lab14}. At the initial phase of ILC with collision energy of 500 GeV, the double Higgs Strahlung mechanism is employed to find out the triple Higgs coupling. While at the second phase of the ILC or CLIC with collision energy at a multi-TeV scale, the W fusion will be significant as the cross section increases with the collision energy due tot-channel enhancement \cite{lab15, lab16}. For double Higgs boson generation $e^+ e^- \rightarrow hh^+ X$, the essential criteria being electron-positron collision energy of $\sqrt{s}$ must be greater than 250 GeV. Some applicable Feynman diagrams are shown in figure \ref{fig:1} and \ref{fig:2}. Those in figure \ref{fig:1} are for both THDM and SM whereas figure \ref{fig:2} is only for THDM.\\

The Mandelstam variables \cite{lab17} s,t,u can be explained as:
\begin{itemize}
\item $s=(p_1+ p_2)^2=(p_3+p_4)^2$
\item $t=(p_1-p_3)^2 = (p_4-p_2)^2$
\item  $t=(p_1-p_4)^2 = (p_3-p_2)^2$
\end{itemize}
here $p_1$, $p_2$ and $p_3$, $p_4$ represent the four-momenta of incoming and out going particles, s and t are square of center of mass energy and four momentum transfer respectively. These channels are used to represent various scattering events in which the exchanged intermediate particle's squared four-momenta equals to s, t, u respectively \cite{lab17}. The s-channel represents the process in which particle 1 and 2 join to form an intermediate particle and finally split into particles 3 and 4. It is the only process from where after resonance new unstable particles could be observed. The t-channel is the process where the particle 1 becomes the particle 3 after emitting intermediate particle, whereas the particle 2 becomes 4 after absorbing the intermediate particle .\\ 

\begin{figure}[h]
    \centering
    \includegraphics[width=10cm]{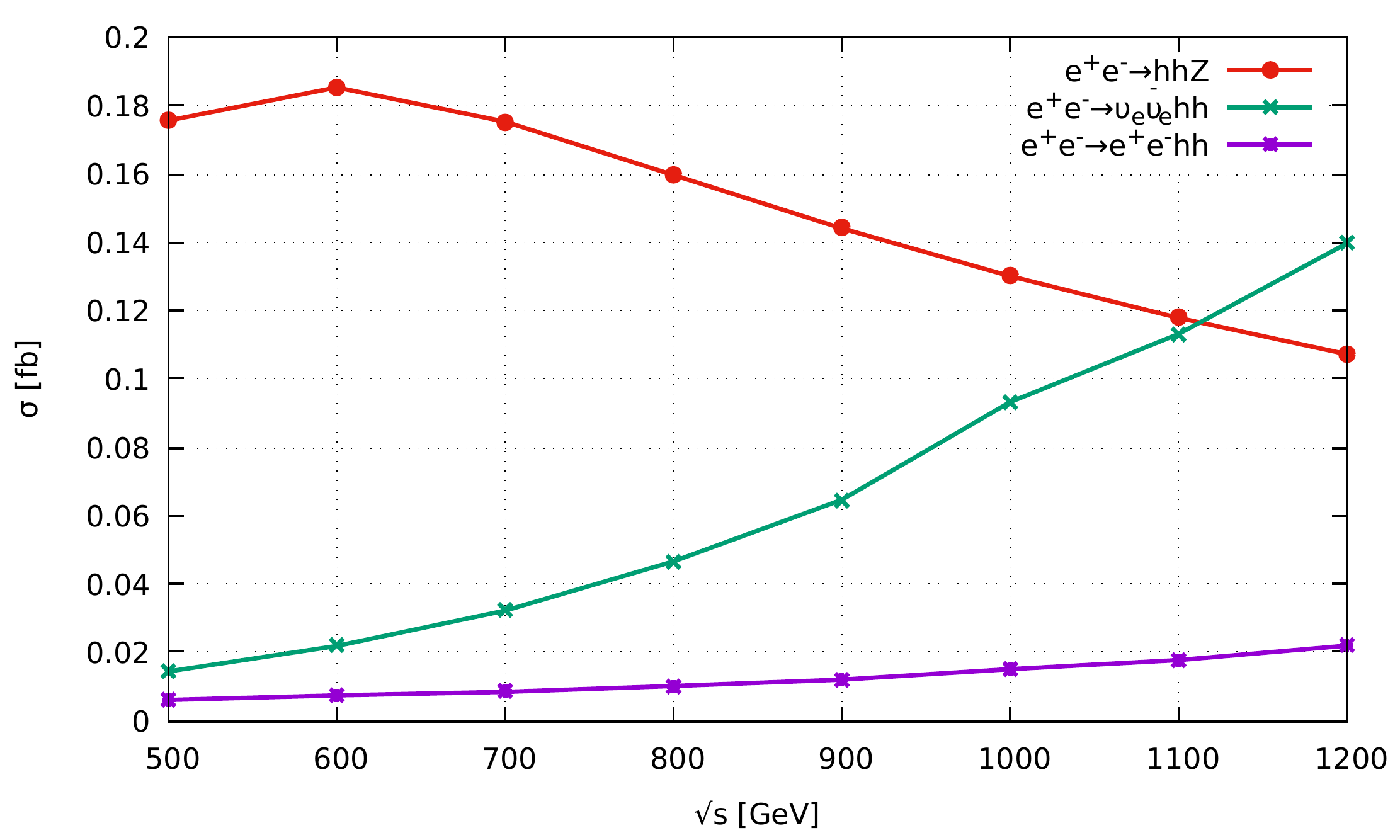}
    \caption{Cross section versus $\sqrt{s}$ for different processes in the SM.}
    \label{fig:3}
\end{figure}

\begin{figure}[h]
    \centering
    \includegraphics[width=10cm]{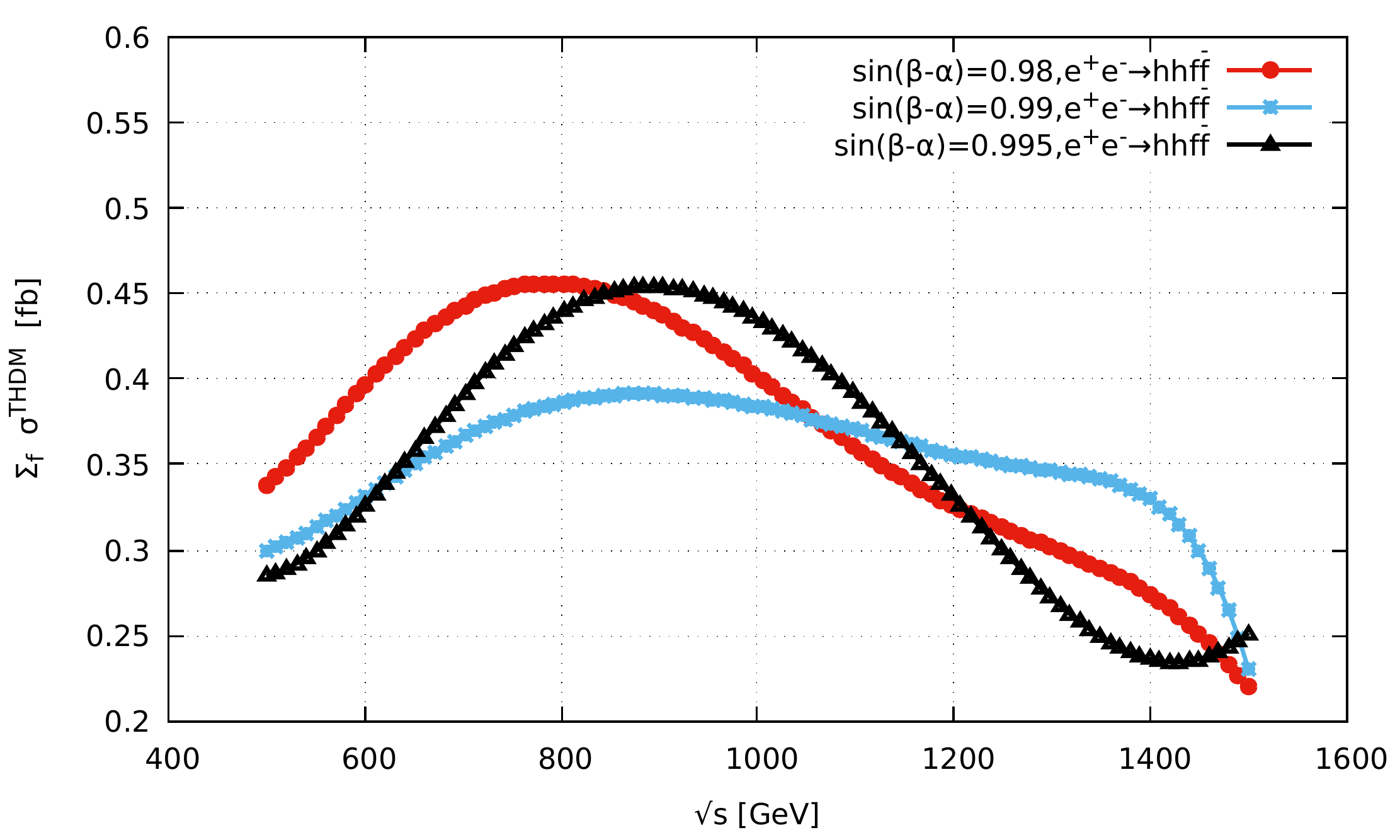}
    \caption{Sum of Cross section versus $\sqrt{s}$ for third generation fermions in the THDM at $s_{\beta-\alpha}= 0.98,0.99$ and $0.995$.}
    \label{fig:4}
\end{figure}

The Double Higgs production in SM contributed via s-channel $e^+e^- \rightarrow hhZ$ shown in diagrams (a)-(e) in Fig. \ref{fig:1} as well as via processes $e^+e^- \rightarrow e^+e^- hh$ and $e^+e^- \rightarrow \nu_e \overline{\nu}_e hh$ shown in diagrams (d)-(f) in Fig. \ref{fig:1}. \ifx Two kinds of diagrams contribute in the double Higgs boson generation process in the SM, one is called s-channel $e^+e^- \rightarrow hhZ$ appeared in diagrams (a)-(e) in Fig. \ref{fig:1}, other are the vector boson fusion processes $e^+e^- \rightarrow e^+e^- hh$ and $e^+e^- \rightarrow \nu_e \over-line{\nu}_e hh$ appeared in diagrams (d)-(f) in Fig. \ref{fig:1}.\fi Their cross section values which are obtained are written in the table \ref{table:3} and the associated results are drawn in figure \ref{fig:3}. So, it is clear from the above table and plots that s-channel generally determines the production of double Higgs cross section at $\sqrt{s}=500GeV$, but with increase in collision energy there is a decrease in cross section and W fusion mode becomes dominant. Therefore, at the first stage of the ILC \cite{lab7}, $e^+e^- \rightarrow hhZ, (Z \rightarrow f \overline{f})$ mode is sizable in computing triple Higgs boson self coupling. The cross section values for third generation fermions ($f \neq t$) with center of mass energy are also calculated for THDM at $s_{\beta-\alpha}=0.98,0.99$ and $0.995$. Their cross section values which are obtained are written in the table from \ref{table:4} to \ref{table:6} and the associated results are drawn in figure \ref{fig:4}.\\
\begin{table}[h!]
\begin{center}
    \begin{tabular}{|c|c|c|c|c|}
\hline
\textbf{Sr.No} &$\sqrt{s}$ &$\sigma(e^+e^- \rightarrow hhZ)$ & $\sigma(e^+e^- \rightarrow \nu_e \overline{\nu}_{e} hh)$  & $\sigma(e^+e^- \rightarrow e^+e^-hh)$ \\

  & \textbf{GeV} & \textbf{[fb]} & \textbf{[fb]}&\textbf{[fb]}\\
\hline
1 & 500 & 0.1756$\pm$0.000018 & 0.01452$\pm$0.000084 & 0.00618$\pm$0.000037 \\
2 & 600 & 0.1853$\pm$0.000021 & 0.02197$\pm$0.000174 & 0.00751$\pm$0.000050 \\
3 & 700 & 0.1753$\pm$0.000021 & 0.03229$\pm$0.000442 & 0.00855$\pm$0.000079 \\
4 & 800 & 0.1597$\pm$0.000021 & 0.04667$\pm$0.000695 & 0.01025$\pm$0.000115 \\
5 & 900 & 0.1441$\pm$0.000021 & 0.06472$\pm$0.001119 & 0.01213$\pm$0.000115 \\
6 & 1000 & 0.1301$\pm$0.000021 & 0.09328$\pm$0.007527 & 0.01519$\pm$0.000328 \\
7 & 1100 & 0.1179$\pm$0.000020 & 0.11321$\pm$0.003000 & 0.01783$\pm$0.000326 \\
8 & 1200 & 0.1073$\pm$0.000021 & 0.13972$\pm$0.004694 & 0.02205$\pm$0.000760 \\
\hline
\end{tabular}%
\caption{ Production cross section values and their errors for different processes in the SM.}
\label{table:3}
\end{center}
\end{table}

It has been observed from the above plots that in THDM there is also decrement in cross section
with increase in the collision energy for s-channel process. Therefore, from both figures \ref{fig:3} and \ref{fig:4} it is noticed that at the   begining of ILC \cite{lab7}, $e+ e^- \rightarrow hhZ, (Z\rightarrow f\overline{f})$ mode is sizable in computing triple Higgs boson self-coupling in SM as well as in THDM.\\

\subsection{Double Higgs Boson Production Cross section in THDM}
It may be seen from Fig. \ref{fig:2} that in THDM additional neutral Higgs boson H or pseudo scalar A is present in the figures in the similar manner in Fig. \ref{fig:1}. To check how these diagrams contribute figure (A) in Fig. \ref{fig:2} is considered. For H to be on-shell, its mass must be $250GeV \leq m_H \leq \sqrt{s}-m_Z$. So, the cross section of the diagram (A) can be calculated by multiplying cross section of two-body $e^+e^- \rightarrow ZH$ and branching ratio of $H \rightarrow hh$ decay by taking assuming $\Gamma_H << m_H$, where $\Gamma_H$ represents the total width of H. So, the factor $16 {\pi^2}$ X $c^{2}_{\beta-\alpha}$ X BR $(H \rightarrow hh)$ is multiplied to acquire cross section as compared with the the diagram (a) in the Standard Model. Here $16 {\pi^2}$ is because of the proportion of the two-body and three-body phase-space factors and $c^{2}_{\beta-\alpha}$ appears due to normalized coupling of HZZ to the hZZ in the SM. For instance, at $s_{\beta-\alpha}= 0.98$ the above factor takes value $16 {\pi^2}$ X $c^{2}_{\beta-\alpha}= 6.2534$ by presuming BR$(H \rightarrow hh)=1$, at $s_{\beta-\alpha}=0.99$, $16 {\pi^2}$ X $c^{2}_{\beta-\alpha}=3.14$ and at $s_{\beta-\alpha}=0.995$ $16 {\pi^2}$ X $c^{2}_{\beta-\alpha}=1.56$. Similarly on considering the diagram (B), the same enhancement can be acquired when A is on-shell. Therefore, this shows that the double Higgs boson total cross section may be some times greater as compared to the SM prediction. It is noticeable that this modification in cross section appears upon leaving the alignment limit that is $s_{\beta-\alpha} \neq 1$ as both HZZ coupling and AZh coupling both are proportionate to $c_{\beta-\alpha}$. So, the enhancement and the divergence in the Higgs boson couplings are strongly correlated with one another as predicted by the SM.

\section{Numerical Results and Discussion}
\subsection{Cross section and Scaling Factor Calculation}
The aim of this study is to compute the cross section of double Higgs boson generation process using type-I THDM and then comparing with the SM cross section values. Finally a convincing correlation has found among the cross section enhancement factor  and scaling factor of the $hf\overline{f}$ couplings. In present case the cross section of the $ e^+e^- \rightarrow hhf\overline{f}$ process where $(f \neq t)$ is calculated. For simplicity only third generation fermions (bottom quark, tau lepton, tau neutrino) are considered for cross section calculation, as they have dominant contribution. All calculations have been performed at $\sqrt{s}=500GeV$ and $1 \leq \tan\beta \leq 30 $. Here without the alignment limit case is considered i.e., $s_{\beta-\alpha} \neq 1$ for $m_H = m_A = m_{H^{\pm}} = 300 GeV$ and $m_H = 300 GeV$, $m_A = m_{H^{\pm}} = 500 GeV$.

\subsection{Scaling Factor for the Yukawa Coupling}
It is the ratio of the Yukawa coupling $ g_{hff}^{THDM}$ in the THDM to the SM value $g_{hff}^{SM}$ \cite{lab8} and can be explained as:
\begin{equation}
 k_f \equiv \frac{g_{hff}^{THDM}}{g_{hff}^{SM}}=s_{\beta-\alpha} + c_{\beta-\alpha} \cot\beta \label{eq:5.1}
 \end{equation}
Due to type-I THDM, the scaling factor $k_f$  is independent on the selection of a fermion f. Equation (\ref{eq:5.1}) shows that we can obtain $k_f < s_{\beta-\alpha} (k_f > s_{\beta-\alpha})$ by considering the sign of $c_{\beta-\alpha}$ to be negative (positive) and $k_f = s_{\beta-\alpha}$ for the limit of $\tan\beta \rightarrow \infty$ \cite{lab8}. Masses $m_H = m_A = m_{H^{\pm}} = 300 GeV$ and $m_H = 300 GeV$, $m_A = m_{H^{\pm}} = 500 GeV$ are considered to calculate $k_f$  for $1 \leq \tan\beta \leq 30 $ values. 

\subsection{Enhancement Factor of the Cross Section}

\begin{figure}
\centering
\begin{minipage}[t]{220pt}
  \centering
  \includegraphics[height=150pt, width=210pt, valign=t]{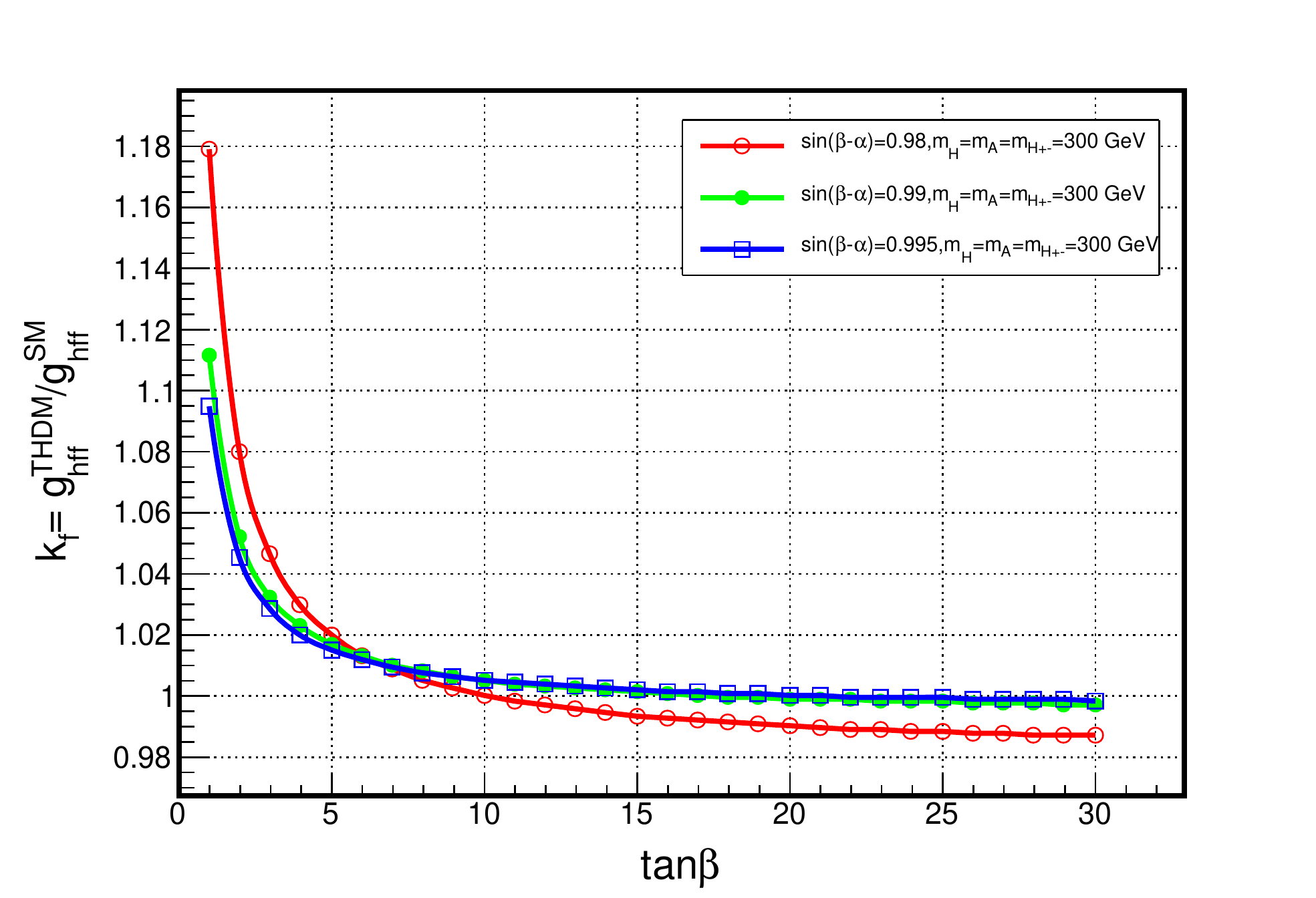}
  \caption{Plot between the scaling factor and $\tan \beta$ at $m_H=m_A=m_{H^{\pm}}=300$ GeV}
  \label{fig:5}
\end{minipage}%
\begin{minipage}[t]{220pt}
  \centering
  \includegraphics[height=150pt, width=210pt, valign=t]{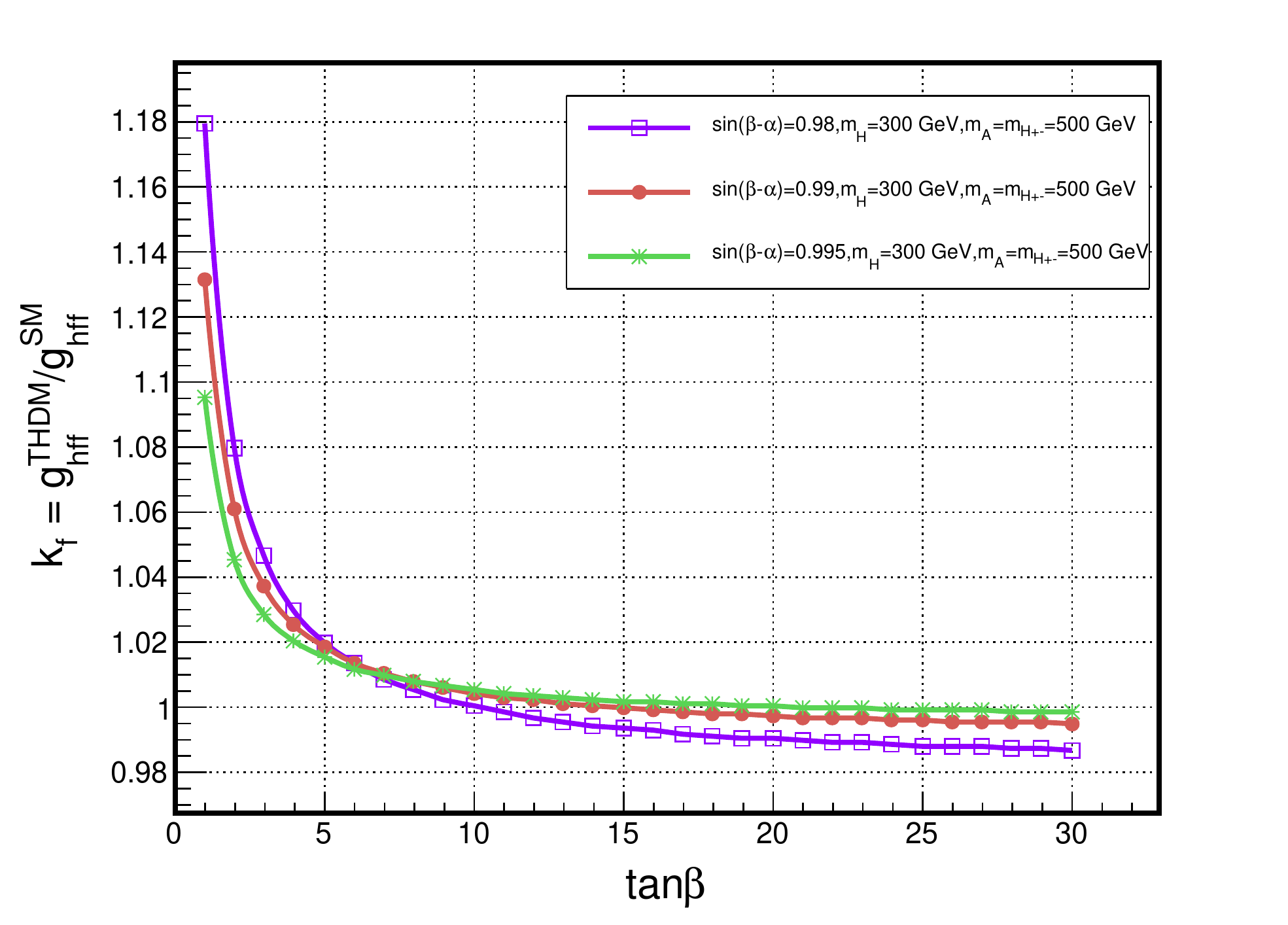}
  \caption{Plot b/w the scaling factor and $\tan \beta$ for large masses $m_A=m_{H^{\pm}}=500$ GeV whereas $m_H=300GeV$}
  \label{fig:6}
\end{minipage}
\end{figure}

 It is defined as the ratio of double Higgs boson production cross section in the type-I THDM to that in the SM \cite{lab8} and is given by:
\begin{equation}
R \equiv \frac{\sum_f \sigma^{THDM} (e^+e^- \rightarrow hhf\overline{f})}{\sum_f \sigma^{SM} (e^+e^- \rightarrow hhf\overline{f})}
\end{equation}
where the summation for f in the above expression is for all fermions except top quark. The sum of cross section for all third generation fermions except top quark in type-I THDM, scaling factor $k_f$ and enhancement factor R plots are given in the figures from figure (\ref{fig:5}) to figure (\ref{fig:10}) respectively. While for the Standard Model (SM) cross section values for third generation fermions except top quark is given in the table (\ref{table:7}).\\

\begin{table}[h!]
\begin{center}
    \begin{tabular}{|c|c|c|}
\hline
\textbf{Sr. No} & \textbf{Process} & \textbf{Cross section} $\bigl<\sigma^{SM} \bigr>$ \\

  &  & \textbf{[fb]} \\
\hline
1& $e^+e^- \rightarrow hhb\overline{b}$&0.025644$\pm$0.0001587 \\
2&$e^+e^- \rightarrow hh \tau\overline{\tau}$ &0.005849$\pm$0.0000364 \\
3&$e^+e^- \rightarrow hh \nu_{\tau}\overline{\nu}_{\tau}$ &0.011592$\pm$0.0000733 \\
\hline
\multicolumn{3}{c}{$\sum_f \sigma^{SM}=0.043085fb$} \\
\end{tabular}%
\caption{The Standard Model (SM) cross section values and their errors for third generation fermions at $\sqrt{s}=500 GeV$.}
 \label{table:7} 
\end{center}
\end{table}

\begin{figure}
\centering
\begin{minipage}[t]{220pt}
  \centering
  \includegraphics[height=150pt, width=210pt, valign=t]{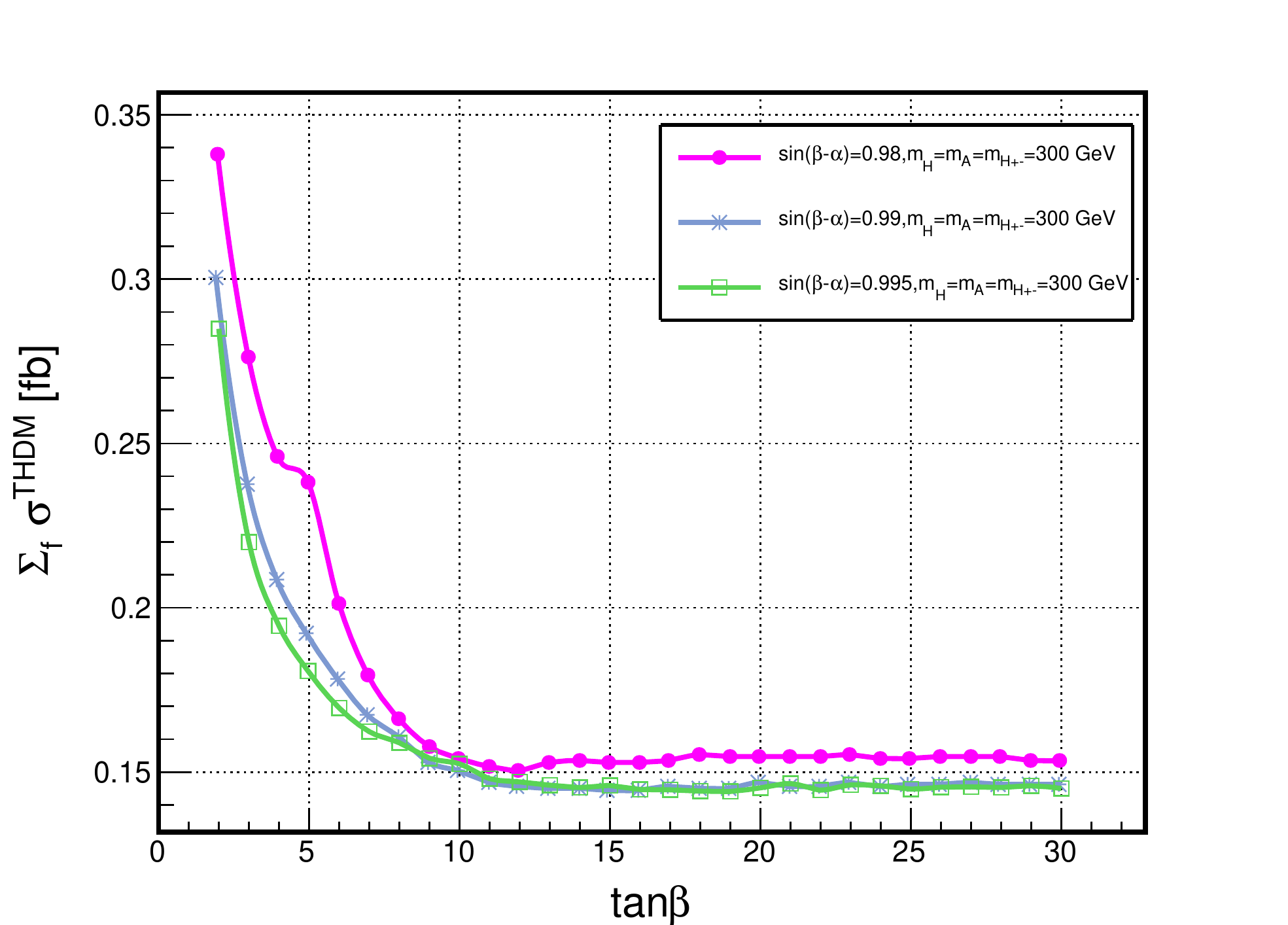}
  \caption{Sum of cross section versus $\tan \beta$ at $m_H=m_A=m_{H^{\pm}}=300GeV$}
  \label{fig:7}
\end{minipage}%
\begin{minipage}[t]{220pt}
  \centering
  \includegraphics[height=150pt, width=210pt, valign=t]{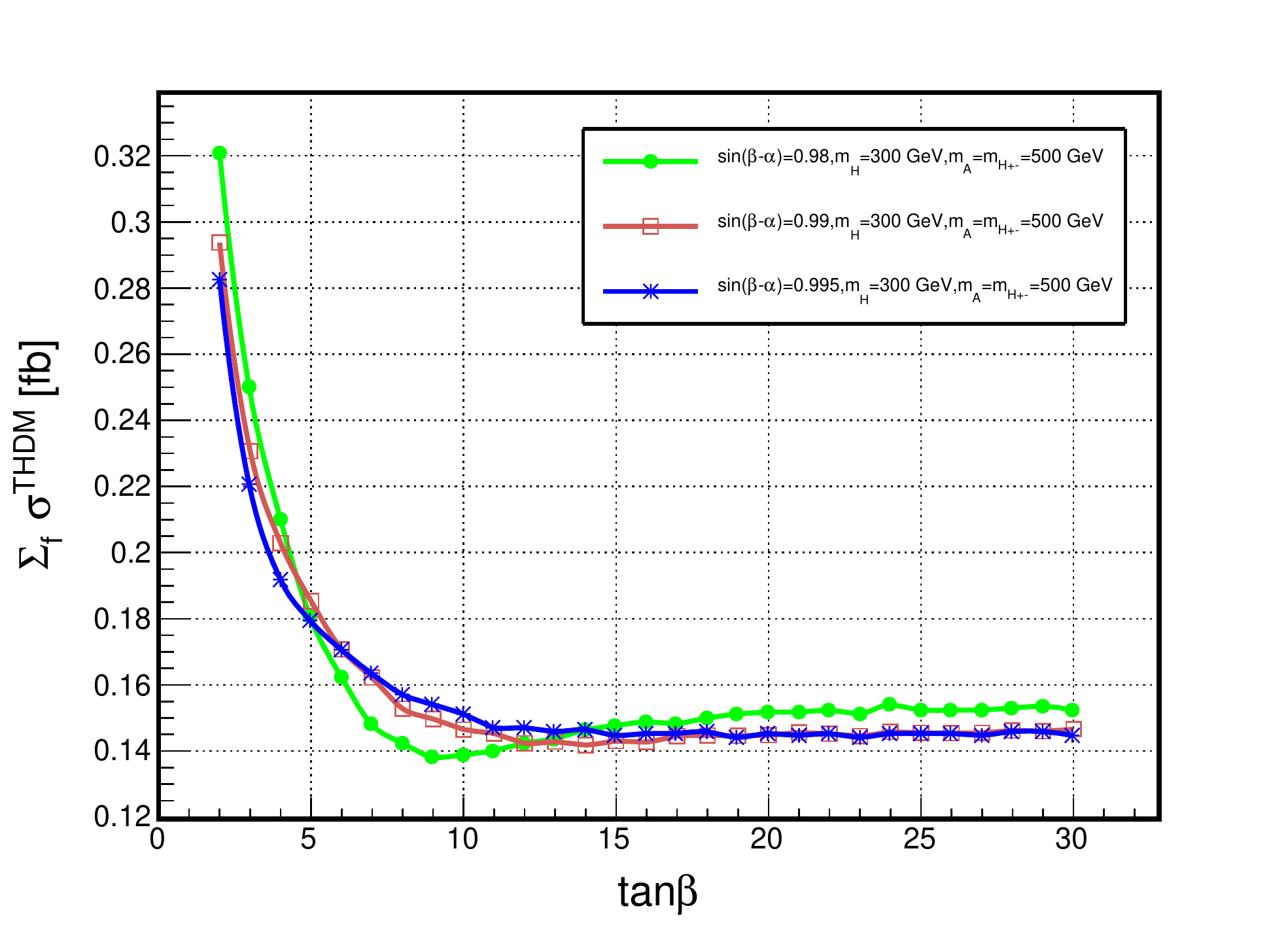}
  \caption{Plot between sum of cross section and $\tan \beta$ for large masses $m_A=m_{H^{\pm}}=500GeV$ whereas $m_H=300GeV$}
  \label{fig:8}
\end{minipage}
\end{figure}

\begin{figure}
\centering
\begin{minipage}[t]{220pt}
  \centering
  \includegraphics[height=150pt, width=210pt, valign=t]{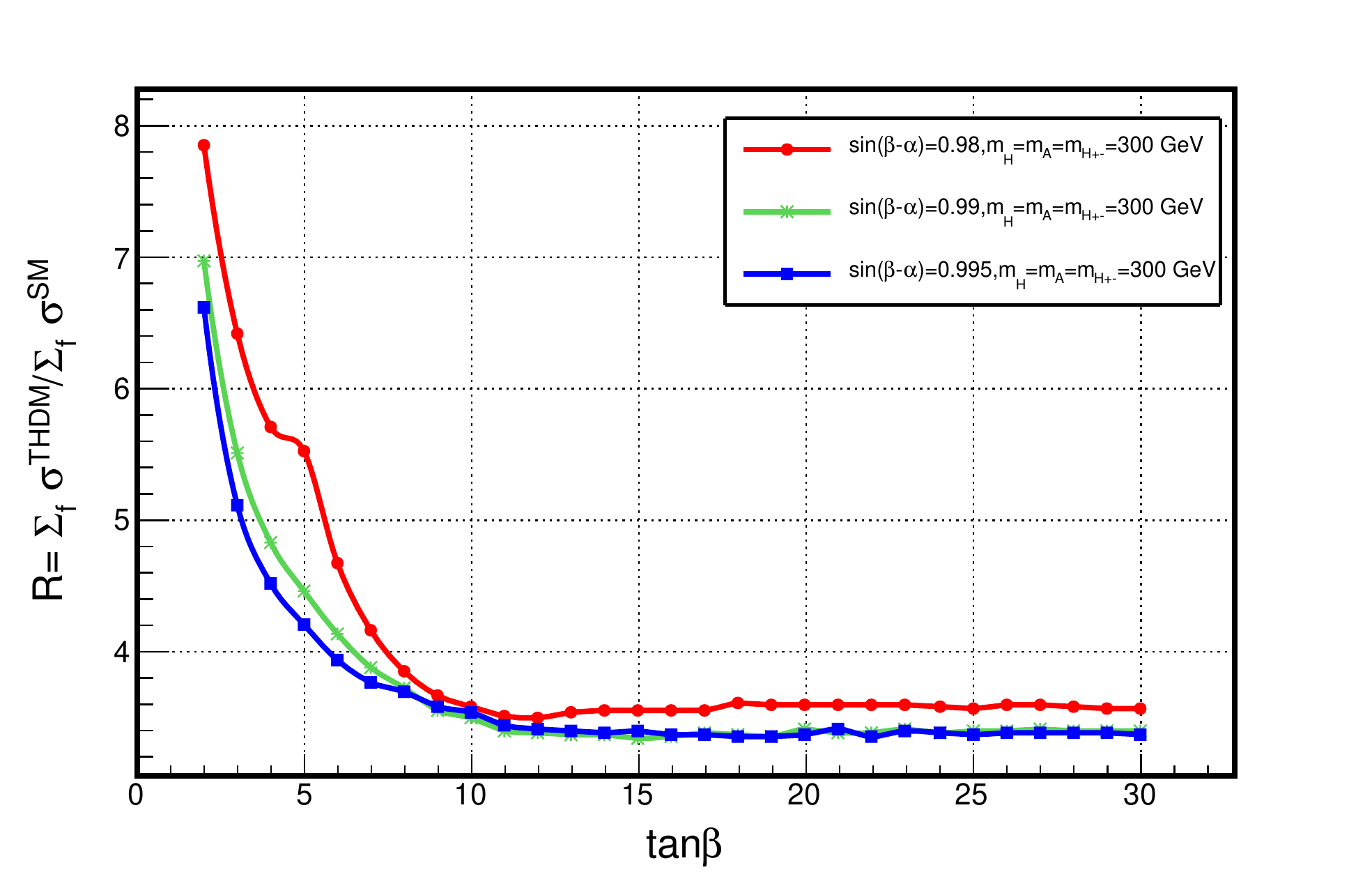}
  \caption{Enhancement factor versus $\tan \beta$ at $m_H=m_A=m_{H^{\pm}}=300GeV$ }
  \label{fig:9}
\end{minipage}%
\begin{minipage}[t]{220pt}
  \centering
  \includegraphics[height=150pt, width=210pt, valign=t]{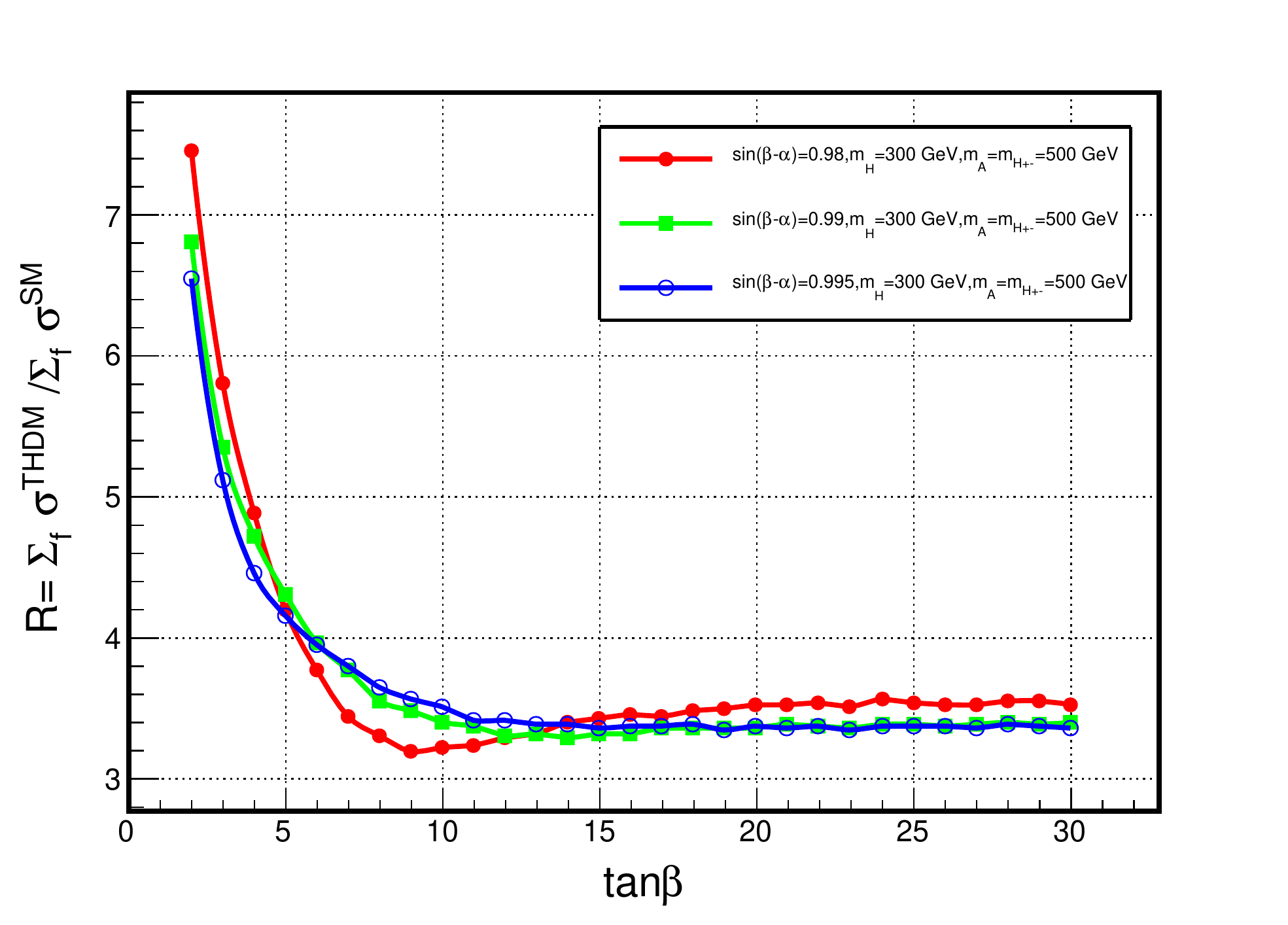}
  \caption{Plot between the enhancement factor and $\tan \beta$ for large masses $m_A=m_{H^{\pm}}=500GeV$ whereas $m_H=300GeV$ }
  \label{fig:10}
\end{minipage}
\end{figure}

\begin{figure}
\centering
\begin{minipage}[t]{220pt}
  \centering
  \includegraphics[height=150pt, width=210pt, valign=t]{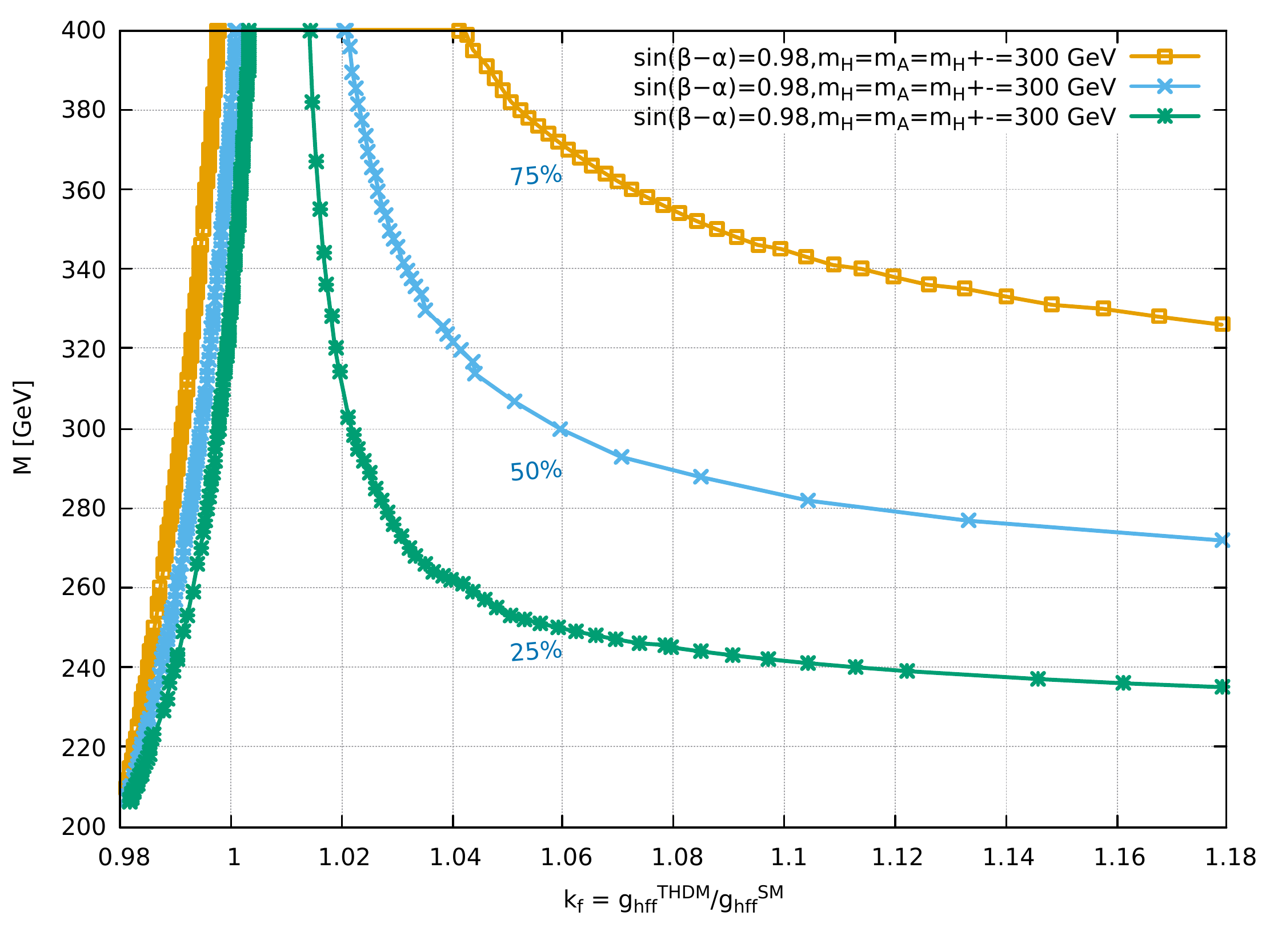}
  \caption{Plot between the scaling factor and M at $m_H=m_A=m_{H^{\pm}}=300 GeV$ and $s_{\beta-\alpha=0.98}$. The green, blue and gold plots respectively display the curves for BR(H $\rightarrow$ hh)=0.25, 0.5 and 0.75}
  \label{fig:11}
\end{minipage}%
\begin{minipage}[t]{220pt}
  \centering
  \includegraphics[height=150pt, width=210pt, valign=t]{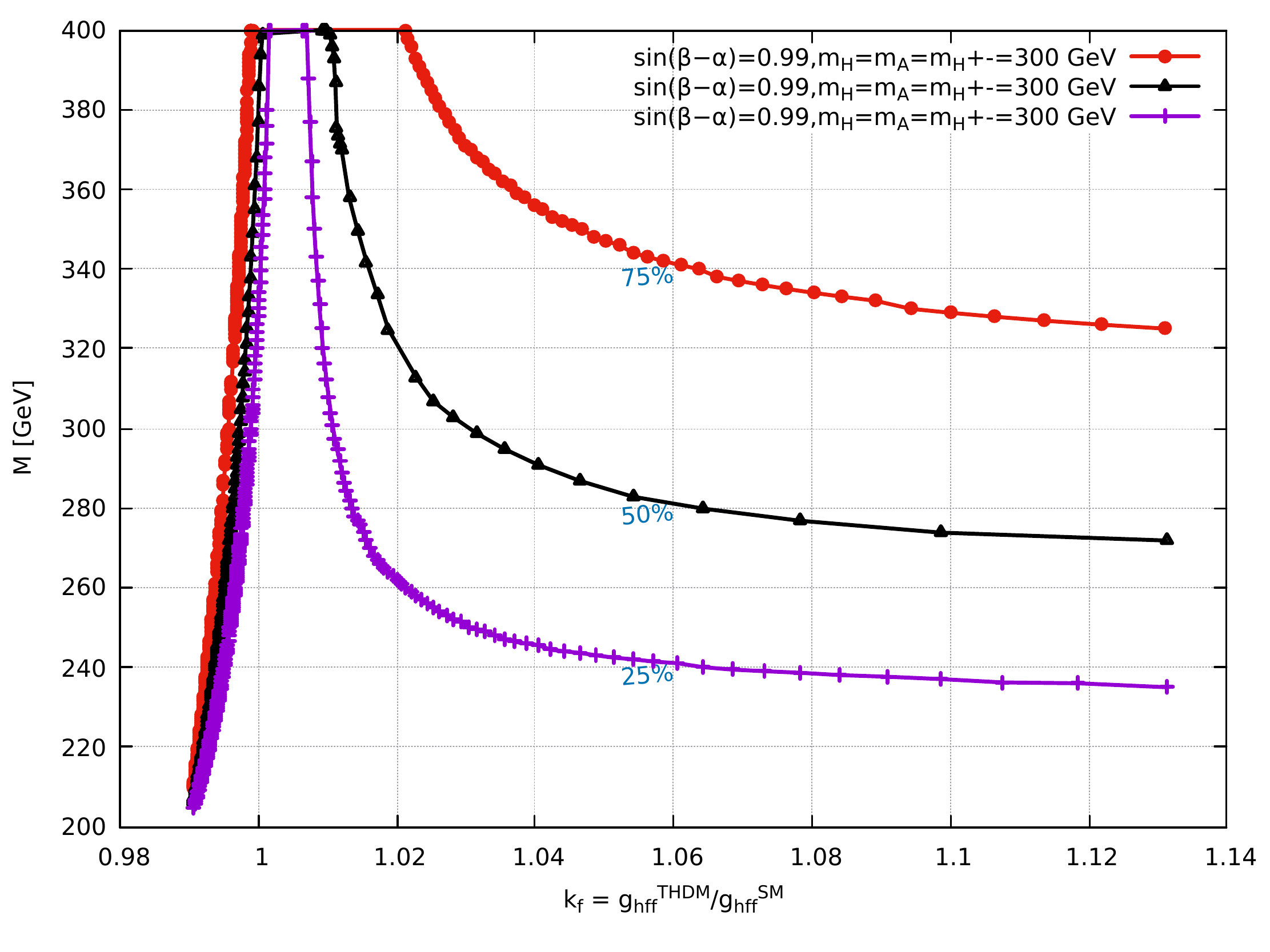}
  \caption{Plot between the scaling factor and M at and $m_H=m_A=m_{H^{\pm}}=300 GeV$ and $s_{\beta-\alpha=0.99}$. The purple, black and red plots respectively display the curves for BR(H $\rightarrow$ hh)=0.25, 0.5 and 0.75}
  \label{fig:12}
\end{minipage}
\end{figure}

\begin{figure}
\centering
\begin{minipage}[t]{220pt}
  \centering
  \includegraphics[height=150pt, width=210pt, valign=t]{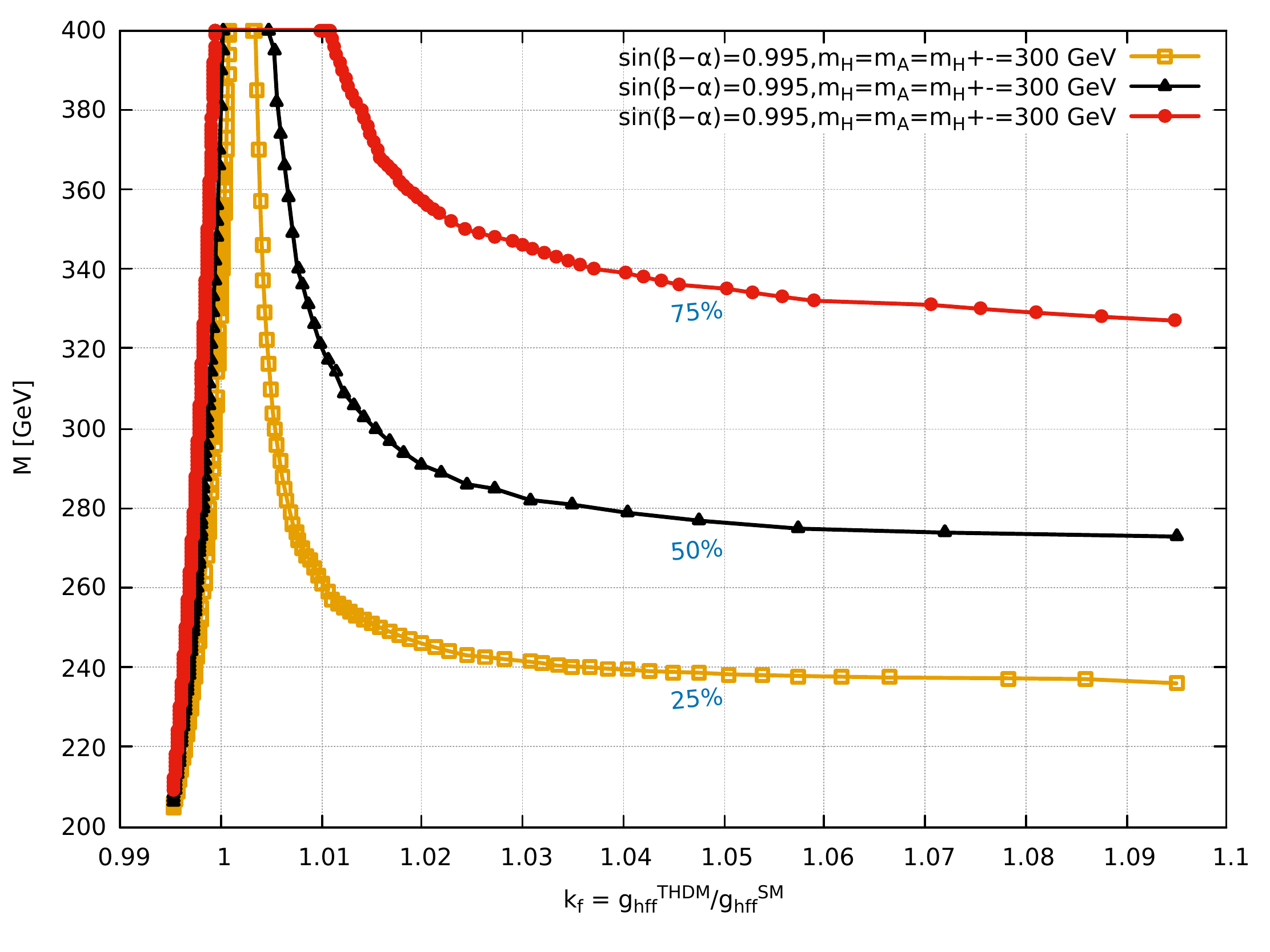}
  \caption{Plot between the scaling factor and M at and $m_H=m_A=m_{H^{\pm}}=300 GeV$ and $s_{\beta-\alpha=0.995}$. The gold, black and red plots respectively display the curves for BR(H $\rightarrow$ hh)=0.25, 0.5 and 0.75}
  \label{fig:13}
\end{minipage}%
\begin{minipage}[t]{220pt}
  \centering
  \includegraphics[height=150pt, width=210pt, valign=t]{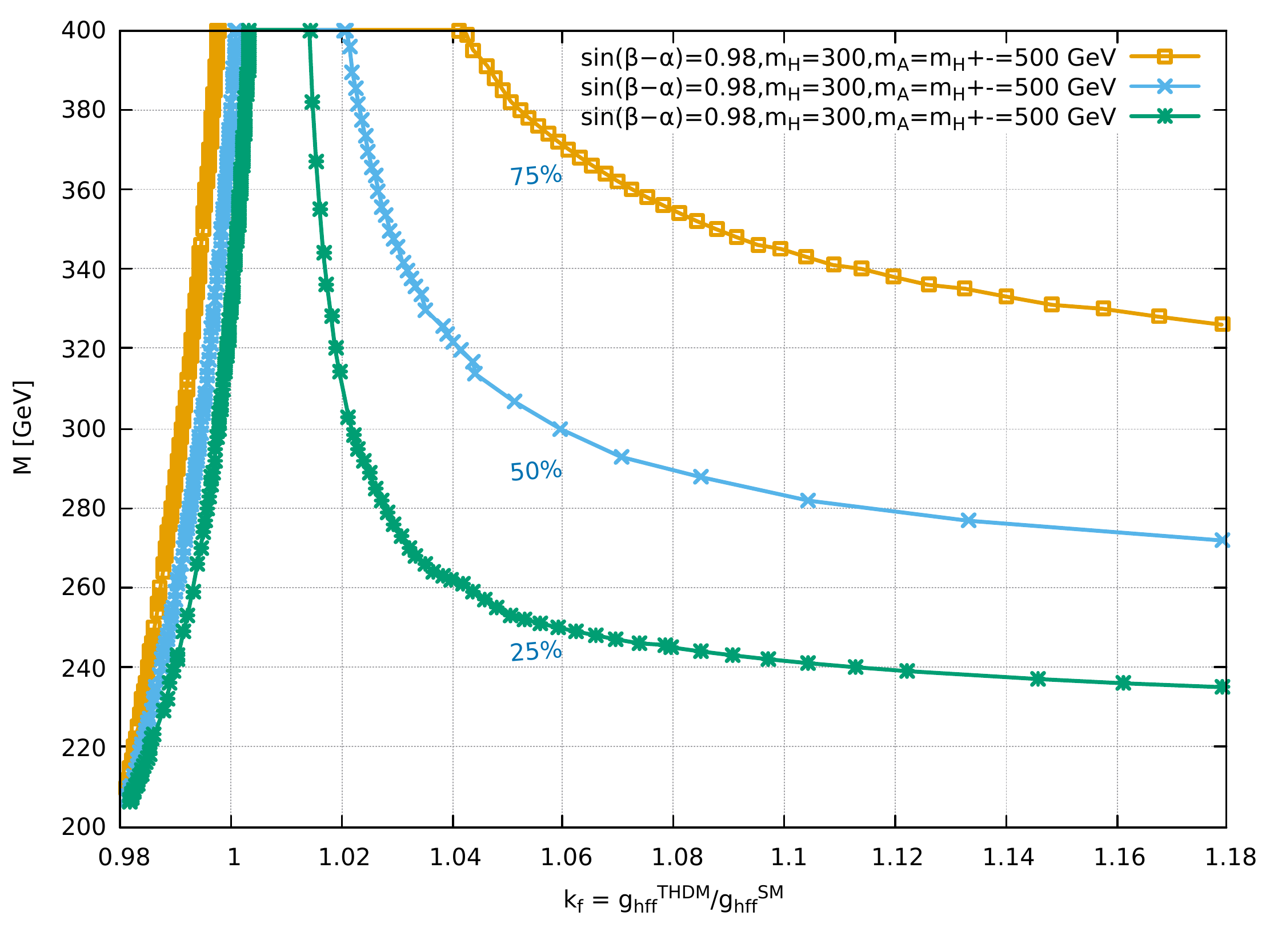}
  \caption{Same as in Fig. \ref{fig:11} but for the case of larger masses $ m_A=m_{H^{\pm}} = 500 GeV$ whereas $m_H = 300 GeV$}
  \label{fig:14}
\end{minipage}
\end{figure}

From all of the tables and figures given above it is clear that in type-I THDM cross section decreases with increase in $\tan\beta$ values. Plot between the coupling factor and $\tan\beta$ shows that for large values of $\tan\beta$, scaling factor becomes equal to $s_{\beta-\alpha}$. Plots in Fig. \ref{fig:6} are similar as in Fig. \ref{fig:5} but in this case $m_A = m_{H^{\pm}} = 500 GeV$ whereas $m_H = 300 GeV$. It is noticeable that cross section value is larger at $s_{\beta-\alpha}$ = 0.98 as compared to $s_{\beta-\alpha}$ = 0.99, 0.995 respectively.\\
This shows that enhancement in cross section occurs on leaving the alignment i.e., $s_{\beta-\alpha} \neq 1$. It can also be observed that a large value of enhancement factor R is acquired at $s_{\beta-\alpha}$ = 0.98 when contrasted with $s_{\beta-\alpha}$ = 0.99, 0.995. Furthermore, there is a decrease in enhancement factor for the case at $ m_H = 300 GeV$, $m_A = m_{H^{\pm}} = 500 GeV$. Plots between the scaling factor $k_f$ and M in figure from (\ref{fig:11}) to (\ref{fig:13}) display the curves of the branching ratio of the ($H \rightarrow hh$) process that is significant for understanding the reason of enhancement of crosssection during the double higgs bosons production.  They also represent the area of the parameter space precluded by the constraints described in the previous chapter. In these plots masses are $m_H = m_A = m_{H^{\pm}} = 300 GeV$ at $s_{\beta-\alpha}$= 0.98, $s_{\beta-\alpha}$= 0.99 and $s_{\beta-\alpha}$= 0.995 respectively. While the parameters $\tan\beta$ and $M^2$ are examined in these graphs.\\

From these graphs it is noticed that experimental constraints are significant in area having small $\tan\beta$, that is large values of $|1-k_f|$, where the quest at LHC \cite{lab5}, peculiarly for $H \rightarrow ZZ$ has dominant contribution for exclusion. This may be interpreted by the way that the  $gg \rightarrow H$ production cross section is proportionate to $\cot^2 {\beta}$ in the restriction of $s_{\beta-\alpha} \rightarrow 1$, such that the constraint can be avoided at large $\tan\beta$ case because of small cross section.\\
One more fact we may find from this figure is that the theoretical constraints are significant in the region with large  $\tan\beta$ values or large difference between $m^2_H$ and $M^2$. The particular behavior of the discipline does not vary significantly between $s_{\beta-\alpha}$ = 0.99 and $s_{\beta-\alpha}$ = 0.995, but small values of $|1-k_f |$ are precluded by experimental bounds. It is due to the value of $|1-k_f|$ gets smaller at $s_{\beta - \alpha}$ = 0.995 when contrasted with $s_{\beta-\alpha}$ = 0.99 by a similar value of $\tan\beta$ as shown in equation (\ref{eq:5.1}).\\

\begin{figure}
\centering
\begin{minipage}[t]{220pt}
  \centering
  \includegraphics[height=150pt, width=210pt, valign=t]{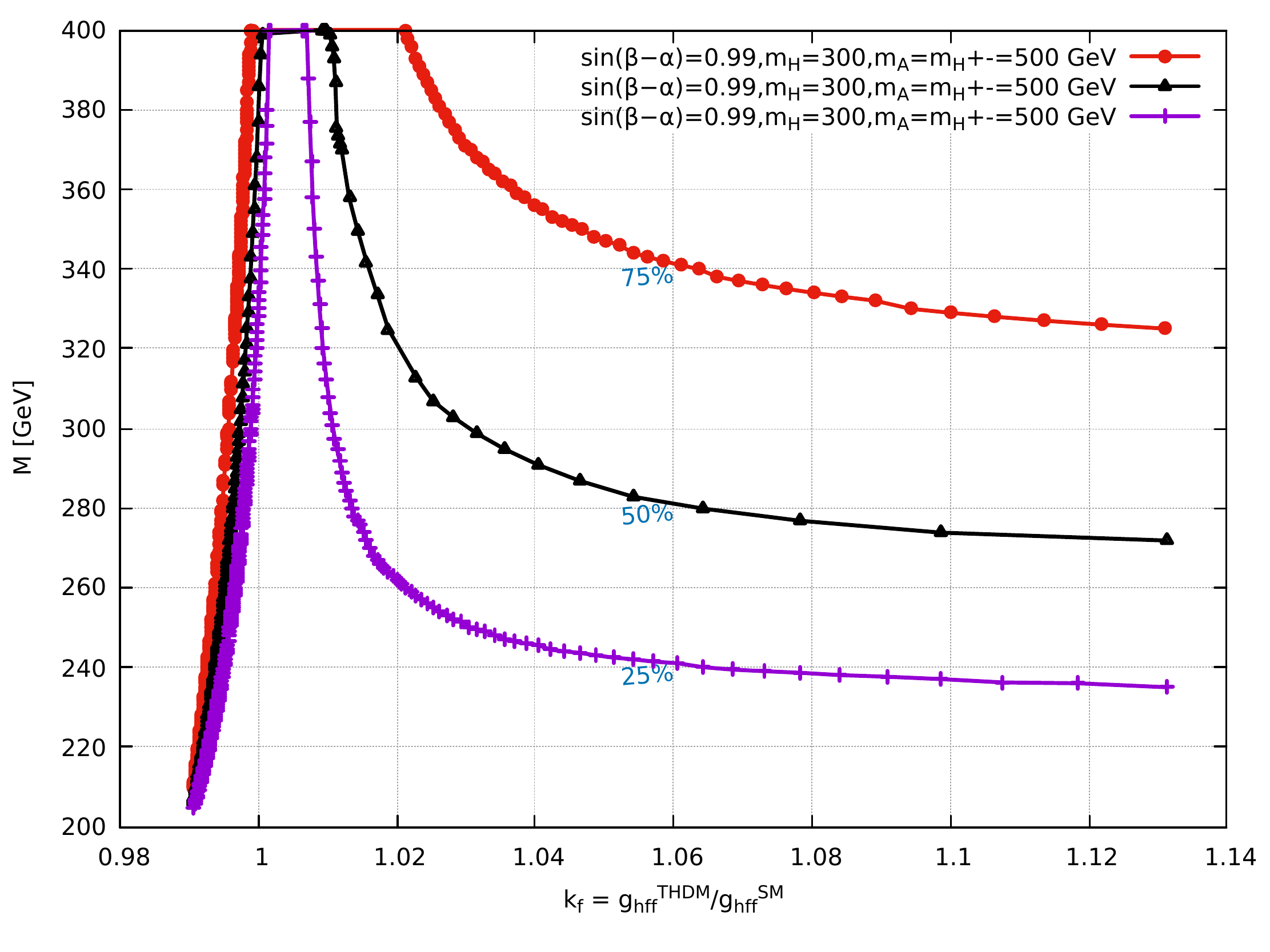}
  \caption{Same as in Fig. \ref{fig:12} but for large masses $ m_A=m_{H^{\pm}} = 500 GeV$ whereas $m_H = 300 GeV$}
  \label{fig:15}
\end{minipage}%
\begin{minipage}[t]{220pt}
  \centering
  \includegraphics[height=150pt, width=210pt, valign=t]{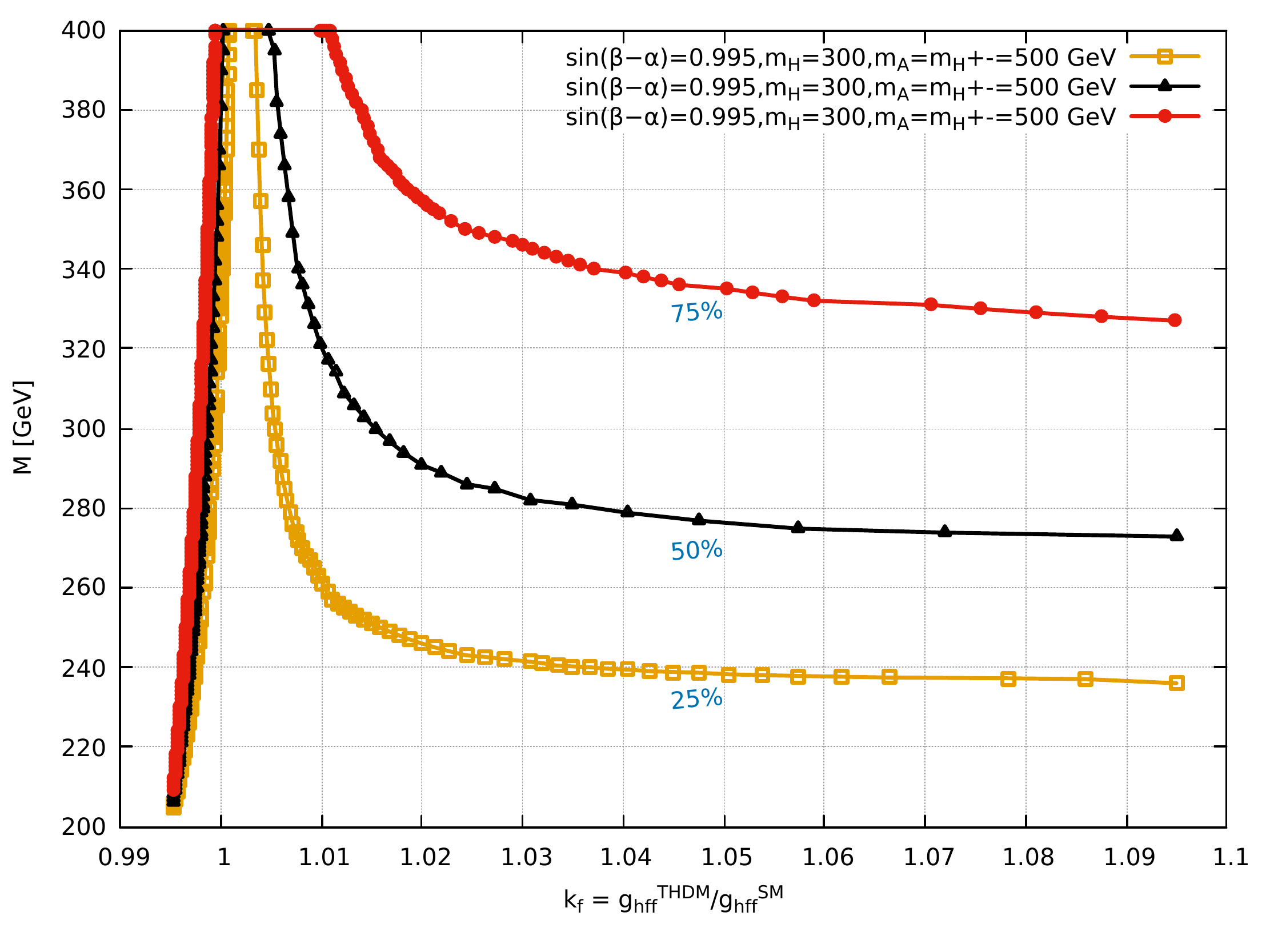}
  \caption{Same as in Fig. \ref{fig:13} but for large masses $ m_A=m_{H^{\pm}} = 500 GeV$ whereas $m_H = 300 GeV$ }
  \label{fig:16}
\end{minipage}
\end{figure}

Similar calculation is performed in figures from (\ref{fig:14}) to (\ref{fig:16}) but for large masses $m_A = m_{H^{\pm}} = 500 GeV$ whereas $m_H = 300 GeV$. The area allowed by the discipline is about the same.\\

\begin{figure}
\centering
\begin{minipage}[t]{220pt}
  \centering
  \includegraphics[height=150pt, width=230pt, valign=t]{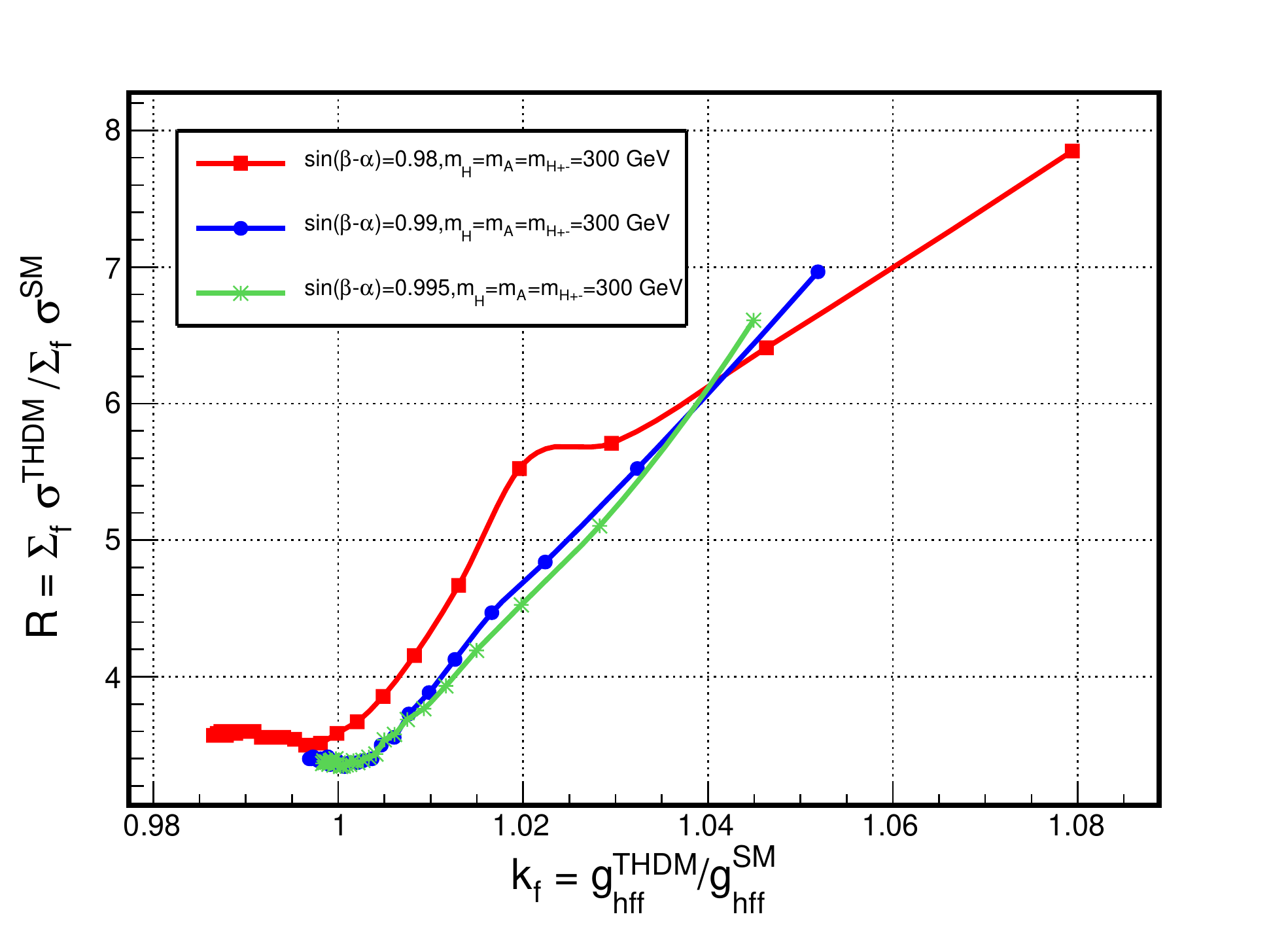}
  \label{fig:15}
\end{minipage}%
\begin{minipage}[t]{220pt}
  \centering
  \includegraphics[height=150pt, width=230pt, valign=t]{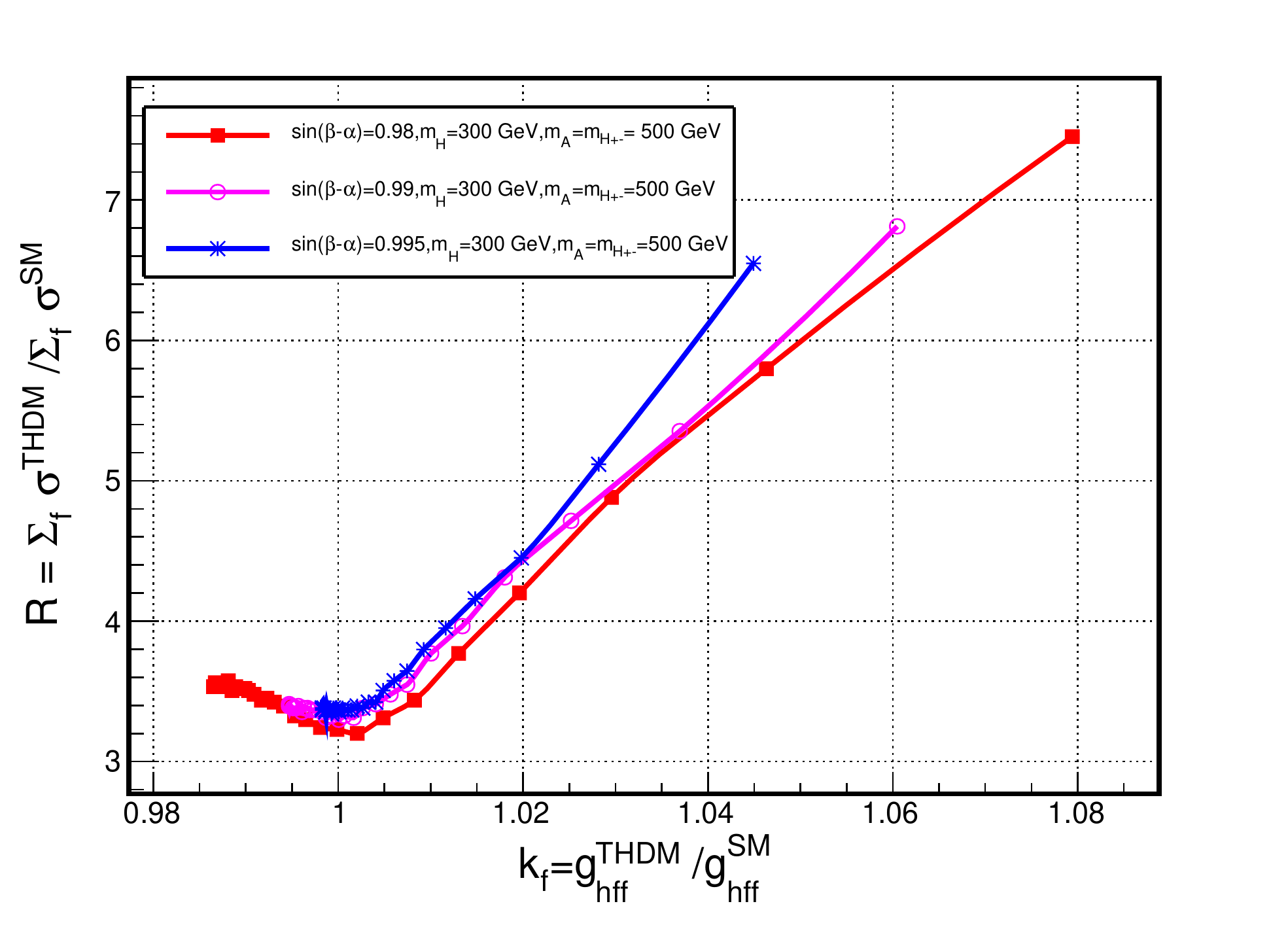}
  \label{fig:16}
\end{minipage}
  \caption{A: Correlation between $k_f$ and R at $m_H = m_A = m_{H^{\pm}}= 300 GeV$. Here $s_{\beta-\alpha}  = 0.98$ (red line), $s_{\beta-\alpha}= 0.99$  (blue line) and $s_{\beta-\alpha}= 0.995$  (green line) B: ame as in Fig. \ref{fig:17} but for large masses $m_A = m_{H^{\pm}}= 500 GeV$ whereas  $m_H = 300 GeV$}
\end{figure}
\ifx
\begin{figure}
  \centering
  \includegraphics[scale=0.34,valign=t]{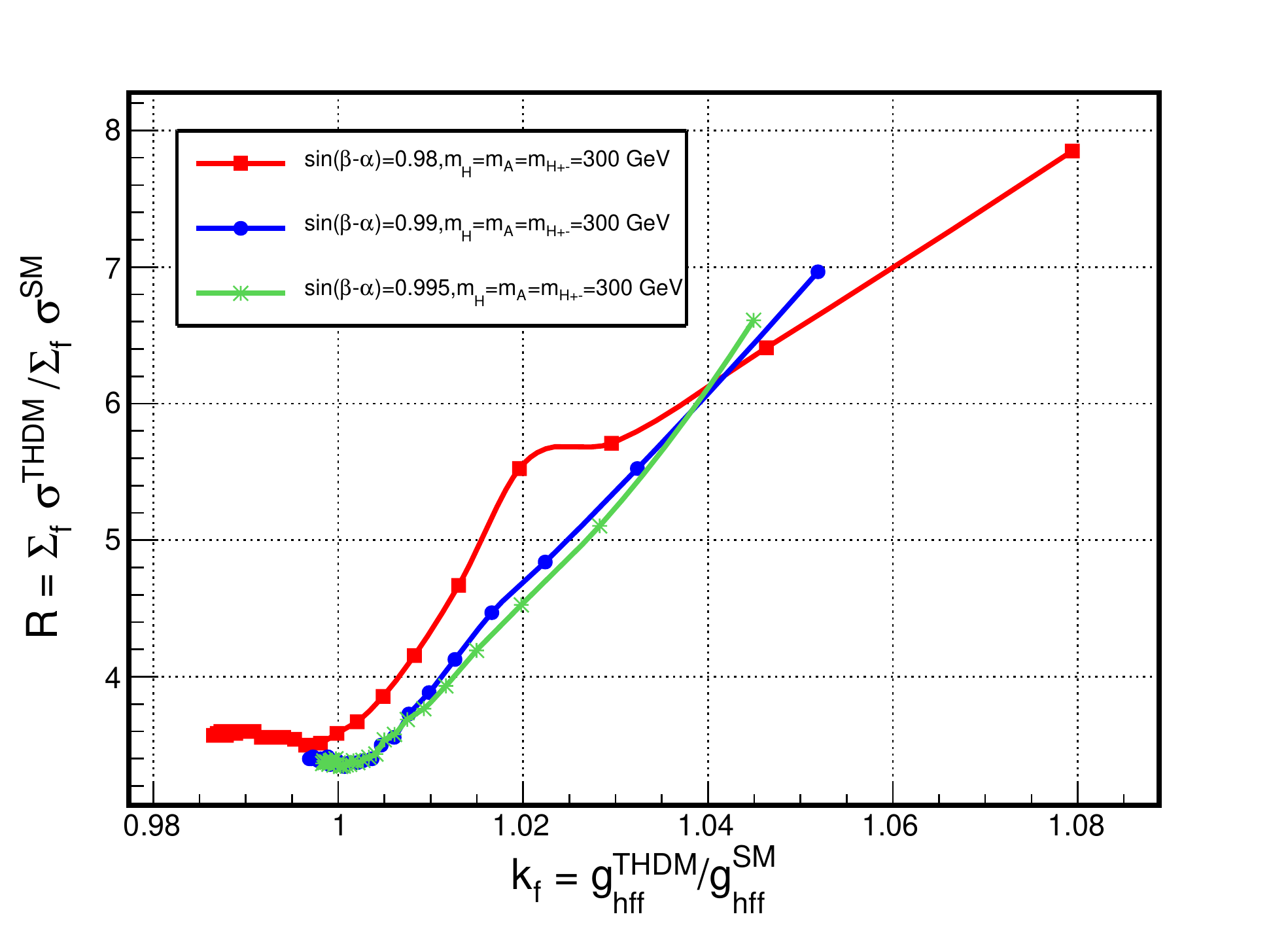}
  \includegraphics[scale=0.34,valign=t]{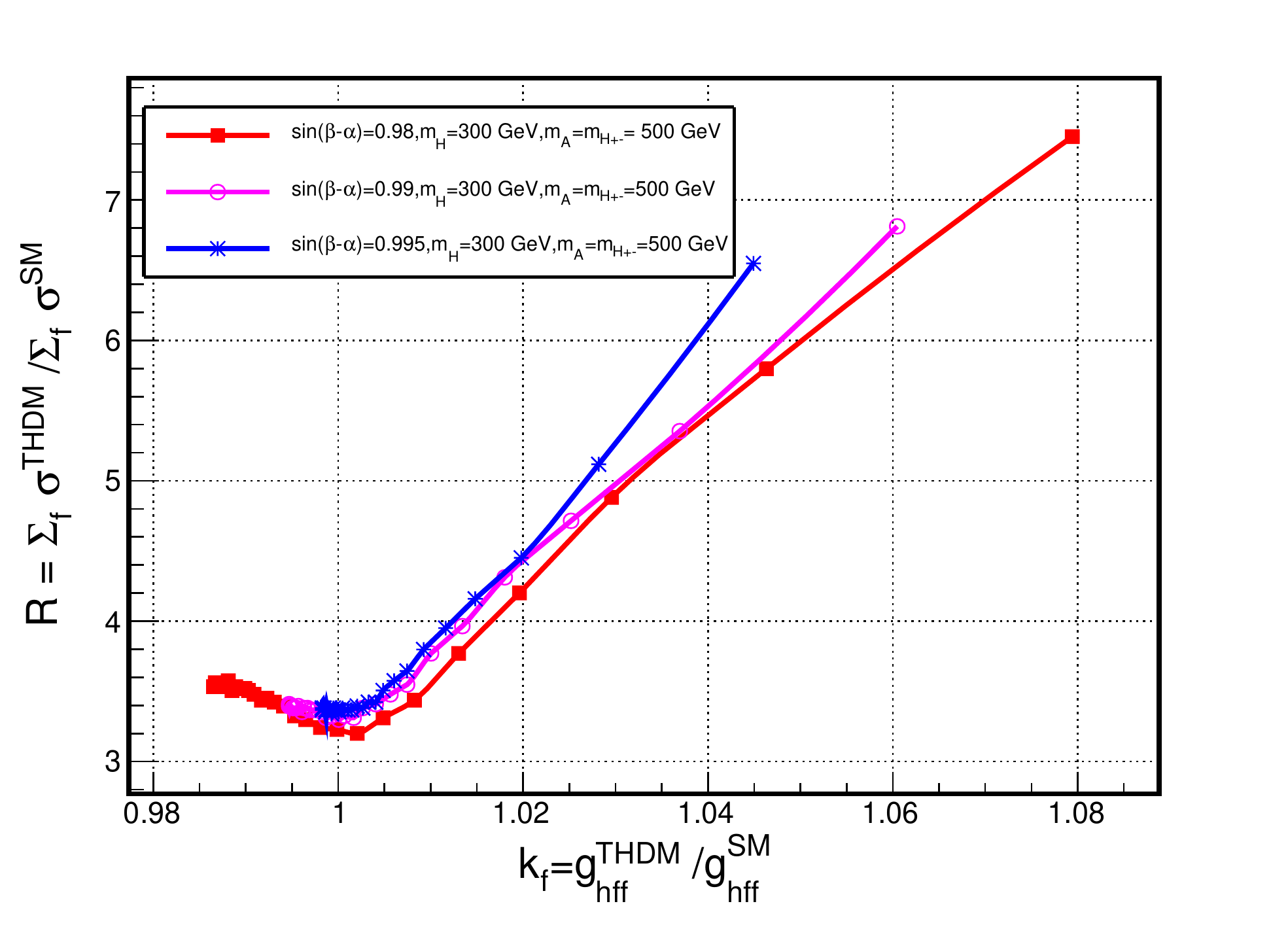}
  \caption{A: Correlation between $k_f$ and R at $m_H = m_A = m_{H^{\pm}}= 300 GeV$. Here $s_{\beta-\alpha}  = 0.98$ (red line), $s_{\beta-\alpha}= 0.99$  (blue line) and $s_{\beta-\alpha}= 0.995$  (green line) B: ame as in Fig. \ref{fig:17} but for large masses $m_A = m_{H^{\pm}}= 500 GeV$ whereas  $m_H = 300 GeV$}
\end{figure}
\fi
The plots in figure (\ref{fig:17}) represent the enhancement factor R as functions of the scaling factor $k_f$. Where red, blue and green line represent $s_{\beta-\alpha}$ = 0.98, 0.99 and 0.995 respectively. Due to the extra neutral Higgs bosons (H \& A) on-shell mediation these plots display a distinct correlation between R and $k_f$ and an appreciable modification of the cross section because of . It is also noticeable that a large value of R is attained at $s_{\beta-\alpha}$ = 0.98 as compared to $s_{\beta-\alpha}$ = 0.99, 0.995, as the AZh and HZZ couplings are proportionate to $c_{\beta-\alpha}$. It can be seen that there is a strong enhancement in the enhanced $\tan\beta$ region. This provides an appreciable deflection $R >1$ even at $k_f$ = 1. It is observed that R value may be about 7.8422 (6.9582, 6.6102) for $s_{\beta-\alpha}$ = 0.98 (0.99, 0.995). The similar calculation is performed in plots shown in Fig. 18 but taking values $m_H = 300 GeV$ and $m_A = m_{H^{\pm}} = 500 GeV$. The region allowed is about same as in Fig. \ref{fig:17}. From this figure it is noticed that pattern of plots are moved below because of the A mediation being off-shell. Here the R value is around 7.4432, 6.8074 and 6.5429 at $s_{\beta-\alpha}$= 0.98, 0.99 and 0.995 respectively.\\

\ifx
 
\begin{figure}[tb]
 \centering
 caption{{Each sample is normalized to the real number of events obtained at 100 $fb^{-1}$}. The charged Higgs is on the top of the total background events at different charged Higgs mass hypotheses independently at tan$\beta = 50$. It shows its visibility for charged Higgs observability. Only dominant backgrounds are labeled.}
   \label{CHmassbins}
 \end{figure}

\fi

\section{Conclusion}
The correlation between the scaling factor of Higgs boson of the SM and the proportion of the cross section for the $e^+ e^- \rightarrow hhf \overline{f} (f \neq t)$ process normalized to the SM prediction in the Type-I 2HDM is demonstrated in this paper. Here without the alignment limit case is considered, that is $s_{\beta-\alpha}$= 0.98, 0.99 and 0.995 in which resonant impacts of the additional neutral Higgs (H \& A) give a considerable modification of the cross section and at tree level the scaling factor value is different from one . It has been observed that by taking into consideration the constraints from vacuum stability, perturbative unitarity, electroweak oblique parameters,compatibility of the signal strengths of the observed Higgs boson and direct searches for heavy Higgs bosons at collider experiments and a sizable modification of the cross section, particularly a few times greater than the SM prediction, may be attained. Its value depends on the masses of additional Higgs bosons and the scaling factor $k_f$ value. It is also observed that at the first stage of ILC the s-channel generally determines the double Higgs boson cross section in SM as well as in THDM and with further increase in collision energy other modes become dominant. The cross section values which are obtained for THDM are ten to fifteen times greater than the SM values. The correlation between the enhancement factor and scaling factor shows that value of enhancement factor R may be about 7.8422 (6.9582, 6.6102) for $s_{\beta-\alpha}$ = 0.98 (0.99,0.995) at $m_H = m_A = m_{H^{\pm}} = 300 GeV$. Whereas the R value is around 7.4432, 6.8074 and 6.5429 at $s_{\beta-\alpha}$ = 0.98, 0.99 and 0.995 respectively for $m_H = 300 GeV$, $m_A = m_{H^{\pm}} = 500 GeV$. It is anticipated to accurately measured the value of scaling factor $k_f$ at future collider experimentation like the high-luminosity LHC and the ILC, particularly with a few percent and one percent level respectively. So, if there exist some deviations in Higgs boson couplings at future colliders, we anticipate the considerable enhancement of the double Higgs boson production and details about the masses of the additional neutral Higgs boson and soft breaking parameter $M^2$ in the Higgs potential can be extracted.\\

\end{document}